\documentstyle[11pt,oldlfont]{article}
\def\Re{\rlap{\rm I}\mkern3mu{\rm R}}
\def\myref#1{(\oldref{#1})}
\let\oldref=\ref
\let\ref=\myref
\def\a{\alpha}
\def\b{\beta}
\def\g{\gamma}
\def\G{\Gamma}
\def\d{\delta}
\def\e{\epsilon}

\def\f{\varphi}
\def\F{\Phi}
\def\k{\kappa}
\def\l{\lambda}
\def\lb{\label}

\def\m{\mu}
\def\n{\nu}
\def\c{\cite}

\def\Si{\Sigma}
\def\r{\rho}

\def\s{\mbox{\boldmath$\sigma$}}
\def\Fb{\mbox{\boldmath$F$}}
\def\gb{\mbox{\boldmath$g$}}
\def\db{\mbox{\boldmath$\delta$}}

\def\st{\mbox{\boldmath$*$}}

\def\S{\mbox{\boldmath$\Sigma$}}
\def\X{\mbox{\boldmath$\Theta$}}
\def\M{\mbox{\boldmath$\Xi$}}
\def\Mb{\mbox{\boldmath$\not \Xi$}}
\def\Ab{\mbox{\boldmath$A$}}
\def\xib{\mbox{\boldmath$\xi$}}
\def\ub{\mbox{\boldmath$\ub$}}
\def\kb{\mbox{\boldmath$\k$}}
\def\eg{\mbox{\boldmath$e$}}
\def\Sg{\mbox{\boldmath$S$}}

\def\t{\mbox{\boldmath$\tau$}}
\def\mb{\mbox{\boldmath$\m$}}
\def\nb{\mbox{\boldmath$\zeta$}}

\def\ub{\mbox{\boldmath$u$}}
\def\p{\partial}

\def\th{\mbox{\boldmath$\theta$}}

\def\w{\mbox{\boldmath$\omega$}}
\def\we{\mbox{\footnotesize \boldmath$\wedge$}}
\def\L{\mbox{\boldmath$L$}}
\def\La{\mbox{\boldmath$\Lambda$}}
\def\J{\mbox{\boldmath$J$}}
\def\U{\mbox{\boldmath$U$}}
\def\Bb{\mbox{\boldmath$B$}}
\def\Gb{\mbox{\boldmath$G$}}
\def\bb{\mbox{\boldmath$\b$}}

\def\i{\mbox{\boldmath$i$}}
\def\R{\mbox{\boldmath$R$}}
\def\li{\mbox{$\mathcal{L}$}}
\def\ex{\mbox{\boldmath$d$}}
\def\cex{\mbox{\boldmath$D$}}

\def\bi{\bibitem}

\def\B{\begin{equation}}
\def\E{\end{equation}}

\def\VS{\vspace{.1in}}

\begin{document}

\thispagestyle{empty}

\begin{flushright}  gr-qc/9804029 \\
                    LPTENS 98/06 \end{flushright}

\vspace*{0.5cm}

\begin{center}{\LARGE { Currents and Superpotentials in classical 
gauge invariant theories I.
Local results with applications to Perfect Fluids and General 
Relativity.}}

\vskip1cm

B.~Julia and S. Silva

\vskip0.2cm

Laboratoire de Physique Th\'eorique CNRS-ENS\\ 
24 rue Lhomond, F-75231 Paris Cedex 05, France
\vskip1.0cm

\begin{minipage}{12cm}\footnotesize

{\bf ABSTRACT}
\bigskip

E. Noether's general analysis of conservation laws has to be completed
 in a Lagrangian theory with local gauge invariance. Bulk
charges are replaced by fluxes at a suitable singularity (in general 
at infinity) of 
so-called superpotentials, namely local functions of the gauge fields
(or more generally of the gauge forms). Some gauge invariant bulk 
charges and current densities
may subsist when distinguished one-dimensional 
subgroups are present.
We shall study mostly local consequences of gauge invariance.
Quite generally there exist local superpotentials analogous
to those of Freud or Bergmann for General Relativity. They are 
parametrized by infinitesimal gauge 
transformations but are afflicted by topological 
ambiguities which one must handle case by case. 
The choice of variational principle: variables, surface terms and
 boundary conditions is crucial.

As a first illustration we propose  a new  {\it  Affine action} that reduces to 
General Relativity upon  
gauge fixing the dilatation (Weyl 1918 like) part of the 
connection and elimination of auxiliary fields. We can also reduce it by 
similar considerations either
to the Palatini action or to the Cartan-Weyl moving frame action and compare 
the associated superpotentials. This illustrates the concept of
 Noether identities. 
We formulate a vanishing theorem for the superpotential and the current
when there is a (Killing) global isometry
or its generalisation. We distinguish between, asymptotic
symmetries and symmetries defined in the bulk.

A second and independent application 
is a geometrical reinterpretation of the convection of vorticity in 
barotropic nonviscous fluids first 
established by Helmholtz-Kelvin, Eckart and Ertel. In the homentropic 
case it can be seen to follow
by a general theorem from the vanishing of the  superpotential
corresponding to the time independent relabelling symmetry. 
The  special diffeomorphism symmetry is, in the absence of dynamical gauge 
field {\it and spin}, associated to a vanishing internal 
transverse momentum flux 
density. We consider also the nonhomentropic case.
We identify the one-dimensional subgroups responsible for the bulk charges
and thus propose an impulsive forcing for creating or destroying selectively 
helicity resp.  enstrophies in odd resp. even dimensions. This is an example of a new and general Forcing Rule.

\bigskip
\end{minipage}
\end{center}
\newpage

\section{Introduction}
\bigskip
\subsection{Generalities}
\bigskip
Eighty years ago E. Noether \c{No} assembled together in a series of 
theorems some consequences of continuous
symmetries of classical actions. Any rigid (Lie) symmetry gives rise 
to a current with the general formula
 \B
\J:=\S -  \delta \phi \we   \frac{\partial \L}{\partial \ex \phi}  \lb{1}
\E
The current is a $(D-1)$-form,
$\delta \phi$ is the variation of the field under an 
infinitesimal rigid symmetry, 
$\ex \S$ represents the total divergence by which the Lagrangian 
density changes (so $\S$ itself 
is defined up to an exact form, in other words up to a topological 
current) and a sum over the independent fields $\phi$ is 
implied. The form notation should not deter the reader as we shall return to
components for the simplest applications. 
 The first theorem of Noether says that the symmetry and some equations 
of motion are encoded as the 
closure of $\J$ modulo the equations of motion. Conversely
such a conserved current  or more precisely the equality of a linear 
combination of the equations of motion to a total divergence implies a rigid 
symmetry. The precise statement eliminates the topological ambiguity in 
the Noether currents,  there is no classical symmetry associated to
a topological current unless one dualizes the theory!

Secondly local (gauge) invariances imply relations between 
the equations of motion, these are now called Noether identities. 
They follow from the triviality of the gauge variations
which reduce the effectiveness of the variational principle.
Conversely the identities imply local invariances. 
In the Hamiltonian formalism time dependent gauge invariance leads to 
primary constraints whereas
secondary constraints follow from space dependent gauge invariance. 
This is now well understood in relativistic theories such as
electromagnetism... 

A third theorem formulated with Hilbert is
in modern language that in the case of gauge invariance, any 
current $ \J_\xi$ 
associated to a one parameter subgroup of generator $\xi$
of the gauge group is equal, modulo the equations of motion, to an 
identically conserved (i.e. topological) local current. 
The idea then was that local invariance destroyed the physical
relevance of the charges of any rigid subgroup.
This claim can be made stronger as was shown in the case of 
General Relativity by Bergmann and his school around 1950 \c{BH,Fl,Bg},
namely there are {\it local} superpotentials $\U$
($(D-2)$-forms) such that on shell, assuming  only fields and their first 
derivatives contribute to the action:   

\B
\J_\xi := \xi . \J + \ex \xi . \we \U \lb{2}
\E

\noindent  is conserved for all $\xi$'s and hence $\J = \ex \U$. Bergmann, 
see \c{Fl} introduced the term strong conservation laws. 
In fact local invariance is still widely and wrongly believed to 
actually prevent the 
existence of any invariant conserved charge, see however the recent \c{BCJ}.  Note that 
the locality of $\U$
does not follow from that of $\J$ even when spacetime is contractible.
If one restricts attention to infinitesimal gauge transformations 
along a fixed generator, one may still  multiply the latter by a
scalar coefficient depending arbitrarily on spacetime coordinates and 
apply the Hilbert-Noether-Bergmann construction to that subalgebra, 
one obtains then
\B
\J_\xi \approx \ex \U_\xi := \ex \left( \xi . \U \right) \lb{3}
\E 

Independently
of these results it was shown in 1981 \c{Ju} that a $p$-form gauge 
invariance corresponding to a  $(p+1)$-form
potential leads to a $(D-p-2)$-form $\J$ that is closed on shell.
In other words
$\ex \J \approx 0$ modulo the equations of motion, generalizing the $p=0$ and 
$p=-1$ cases. If one views Yang-Mills 
invariance as a mixture of $p=0 $ and $p=-1 $ invariances one recovers 
the analog of
Bergmann's analysis. We recognize one half of Maxwell's equations in 
the strong conservation equation
\B
\J \approx \ex \U  =\ex \st \Fb
\E 
In the nonabelian case we may still pick a direction of gauge 
transformations with arbitrary (scalar and $x$ dependent)
magnitude $\xi(x)$ then for this particular abelian subgroup of 
gauge transformations we have the same formula \ref{3}. This is 
the origin of the 't Hooft abelian charges of the dyons, see for
instance \c{AD}.

The discussion of higher conservation laws has been recently 
carefully extended to a generalised Noether theorem 
relating symmetries of various types with generalised charges 
\cite{He} in a cohomological framework.

Now in the case of rigid symmetry, $\J$ is already  afflicted by 
ambiguities, it is well known that they permit the 
constructions  of the symmetrized
or improved energy-momentum tensors. This arbitrariness becomes much 
more serious in the case of gauge invariance
as the ambiguity of the superpotential $\U$ seems to be total.
In fact the litterature on General Relativity is littered with
a host of superpotentials without clear status of respectability. We 
shall concentrate in this paper 
on the local aspects of the theory, in particular on the formulas that 
generalize \ref{1}, their  
dependence on the order of differentiation of the fields and on possible 
surface terms. 

But let us 
recall that the physical measurement of the force leads to 
the value of a gravitational 
mass far from its source by actually assuming asymptotic flatness all 
around it. One could also 
expand around an arbitrary background near infinity and define a mass 
parameter there, this has been carried out in particular for anti de
Sitter asymptotics. In this paper we shall 
focus on the asymptotically flat case.  One puts the laboratory 
at infinite distance from the source(s) in some direction in the sense 
that the metric becomes flat up to order 1/r corrections
then the limit of $\frac{r}{2}(g_{00}+1)$ or $\frac{r}{2}(g_{ii}-1)$ in
 asymptotic rest frame coordinates is the physical mass deviating 
test particles.  The use of arbitrary coordinates requires a geometrical
definition of asymptotics, in other words of the boundary at infinity
(we shall consider spatial infinity in this first paper). We must choose a model
manifold for the neighbourhood of infinity but not its coordinates. Note  that 
this manifold does not have to be close to ours except there.
 A side remark is that local but not global asymptotic flatness would 
force us to distinguish between a local definition  of mass from 
formulas involving total fluxes. Let us take the 
example of electromagnetism and consider an orbifold 
ALM space obtained by 
quotienting $\Re^4$ by $Z_2$ (the sphere at infinity is replaced by 
$\Re P_2$). Clearly if the electric 
field is $\frac{e}{ r^2}$ in Gaussian units its flux is equal to 
$2\pi e$ and not $4\pi e$,
the conical singularity at the origin
affects the relation between the total flux and the local (asymptotic) 
field.  
Similarly total angular momentum perpendicular to the direction of 
the source
is measured by the limit of $\frac{R}{4}\epsilon^{ijk}g_{j0}r_k$
again if one assumes global trivial topology at infinity. We leave 
this global issue for subsequent work.

To the above physical and local definition of charge  
one can compare mathematical formulas, for 
instance the charge may take the
form of a flux at infinity, this is the case for the celebrated ADM 
expression \c{ADM} 
for the total mass of a curved spacetime or the generalisation by 
Regge and Teitelboim \c{RT} in a Hamiltonian description. We shall 
follow here a Lagrangian approach and invert the conventional order of the 
constructions: we shall look for a bulk density such
that its integral is equal to the physical mass given by such 
a flux at infinity. It has not been widely recognized that 
when there is a singularity, even if it is hidden behind an horizon
 and contrary to the abelian case, {\it  bulk} integrals may  
not make physical sense. 

The Nester-Witten form \c{Wi,Ne} can be 
used outside horizons and will be discussed in the next paper of this 
series hereafter called paper II. The proof of the positivity of total ADM 
mass for a general solution with black holes uses
the existence of a supersymmetric extension of the theory it 
localises all the energy outside
the horizons, and  the ``energy density" is positive there.

The special case of a global Killing vector is essentially bringing 
us into an abelian framework as Kaluza-Klein inspired ideas may 
suggest. 
In a general gauge theory we shall call Killing symmetry a Lie algebra 
  generator preserving
the value of the gauge field, for instance isometries of a metric, 
isotropy gauge transformations in Yang-Mills theory, 
Killing spinors for supersymmetry etc... 
The existence of a global (bulk) Killing symmetry leads in general to the 
vanishing of the gauge part of the current density as a generalisation of
the vanishing charge of the photons and of all the Fourier zero
modes of the Kaluza-Klein dimensional reduction. 
In the case of diffeomorphisms  the gauge current may fail to  vanish
because of
a surface term but it does vanish for spatial Killing directions and 
in the vacuum as we shall see.
We shall discuss the general formalism in section 2 
but mostly focus on the rich case of diffeomorphisms.

\bigskip
\subsection{Perfect fluids}
\bigskip
The reader interested in the conservation 
laws of fluids will at this stage be able to skip the 
middle sections (3-5)  on our new formulation of General Relativity and 
should  go directly  to section 6, if he so wishes. Relabelling symmetries 
allow a physically suggestive interpretation of the 
conserved quantities of perfect fluids. These
fluids obey a variational principle involving independent Lagrangian 
coordinates (the labels), the fields are simply the 
Eulerian, or laboratory, coordinates, they admit time independent 
space relabeling gauge invariance without any 
gauge field. In the homentropic (possibly compressible) case the 
relabelings are arbitrary volume preserving diffeomorphisms. 
The corresponding spatial Noether current is purely longitudinal 
because there
is no propagating gauge field in this gauge invariant theory, 
this is the local vorticity conservation in comoving cordinates
\c{Ec}. 
Noether's theorem has been 
invoked before but without superpotentials (see the nice review 
\c{Sa}) and when it was precisely formulated it was
the global theorem that was used as in Taub's description of flows 
(see the review \c{Br}) where the roles of Lagrangian and Eulerian 
variables are exchanged.  It turns out that global (bulk)
conservation laws do exist even in the absence of boundaries. This may 
seem surprising to a field 
theorist; we shall explain this phenomenon and identify the rigid 
symmetries responsible for these charges.  We hope
to return to the effect of boundaries in the future.

We shall also identify a simple mechanism of creation of these 
charges by forcing with an optimum scheme  that could be implemented
numerically and almost experimentally. The problem with experimental 
implementation is not serious, in most (=slow speed) situations the
incompressible approximation is valid and thus one may identify at any chosen
instant Lagrangian and Eulerian coordinates so the forcing mechanism can be 
formulated either theoretically in Lagrangian coordinates or practically of 
course in Eulerian ones. We shall return to the incompressible case in the next paper.
 
This forcing, although impulsive,  is 
reminiscent of the generation of the electric charge of electromagnetic
 dyons
by uniform rotation in internal space \c{TW}   and of geodesic
motion of quasistatic solutions of the variational problem of
magnetic monopole theory \cite{Ma}. 
We shall also explain the relation between homentropic and non homentropic 
situations, in fact a partial breaking from $(D+1)$ to $D$-dimensional 
relabeling symmetry by some marker like the value of the entropy 
changes dramatically the number of local invariants and exchanges the 
properties of even and odd numbers of space-dimensions. 
\bigskip

\subsection{General Relativity}
\bigskip 
The organisation of the rest of the paper is as follows. 
First the notions of 
cascade and abelian cascade  of currents and superpotentials: $J$ or 
$T, U, V...$  are introduced in subsection 2.1 and illustrated  on 
simple examples including electromagnetism, Yang-Mills theory, p-forms
gauge fields, in the rest of section 2. 
The identification of Noether currents for selected generators
reduces the problem of finding the invariant charges to the selection of 
abelian subgroups of gauge symmetries. 

In section 3, Hilbert's action in second order form is analysed
and the need for longer cascades appears.
The spin term of Belinfante's symmetrized energy-momentum tensor \c{Be} for 
matter is derived from the
matter contribution to the superpotential. The mechanism is that 
tensor fields with spin do
transform under diffeomorphisms with derivative terms as gauge fields do.

In the fourth section 
a new first order affine gauge theory of gravitation is defined.
Its symmetries include diffeomorphisms, local linear frame transformations
and a new gauge symmetry without gauge field, let us call it
the Einstein-Weyl symmetry,  we shall see why  momentarily. The latter
gauge symmetry does not have any propagating gauge fields, as a 
consequence the associated superpotential and currents vanish. This is a special case of the
so-called Noether identities which is formulated as a general vanishing theorem in subsection 4.2. The various
superpotentials are easily analysed. In subsection 4.3  other Noether 
identities are discussed and the Sparling-Dubois-Violette-Madore
rewriting of Einstein's equations as a closure, or conservation, condition
\c{Sp,Du} is adapted to our affine theory. Supergravity practitioners 
should not be surprised by such a result,  see for instance \c{Jac}.
What happens here is that the conservation laws encode {\it all}
the equations of motion and not only some combinations of them. Gauge 
invariance far from being a nuisance has the power to determine the dynamics.  

The affine theory  leads, see subsection 4.4,    either to first order
Poincar\' e (Cartan-Weyl) theory by going to orthonormal frames and 
using the metricity of the 
connection or to the Palatini formalism by going to 
a coordinate frame and eliminating the torsion. In both cases one 
eliminates part of the linear 
connection by its equation of motion and by fixing a residual 1-form 
gauge invariance (without gauge 2-form): 
the arbitrariness of the scaling part of the linear connection.
In other words the invariance of the action under the shift of 
$\Gamma^\r_{\m\n}$ by 
a scaling (Weyl) component, $A_\m (x)\ \delta^\r_\n$, is a
gauge symmetry that generalizes the so-called Einstein symmetry 
\c{HeM}. In summary, modulo this ``Einstein-Weyl"  arbitrariness which is
due to the form of the scalar curvature, the vanishing of torsion and 
nonmetricity  follow
from the variations of suitable components of the connection field. 
 The name of H.  Weyl is associated with the invention of (scaling)
gauge invariance and is appropriate despite differences in the implementation.
For this new gauge invariance  one explains again the vanishing 
of the superpotential and hence of the current
by the absence of propagating gauge field. Recall the examples of 
kappa-symmetry  or  (string theory) Weyl currents...      

The local invariance of our affine action with respect 
to those $gl(D,\Re)$ generators
that are not in the Lorentz subalgebra defined by the metric (and 
consequently 
do not propagate) leads also  to the vanishing of the associated 
$U^{(ab)}$ superpotentials. Frauendiener \c{Fr} also considered the full
frame bundle to investigate   energy-momentum pseudotensors.
Finally the Hilbert action follows from our action by going to
second order formalism via the Cartan-Weyl action for instance.

 In Subsection 4.5 we compare the superpotentials associated to
these four actions, they may be called respectively affine, 
Cartan-Weyl, Palatini
and  M\o ller. In order to recover the right mass for the Schwarzschild 
solution M\o ller did 
actually rescale arbitrarily the  potential  derived canonically from 
the Hilbert action  and multiplied  by a factor of two \c{Mo} the honest one. 
A linear combination of the energy-momentum tensor and its associated
superpotential involving an infinitesimal gauge 
parameter gives \ref{3} the ordinary Noether current for the one 
dimensional subgroup 
along  a given gauge direction. Ignoring extra terms due to
higher derivatives it would have the form  
\B
\J_{\xi} =\xi^\r \J_\r +\ex \xi^\r \we \U_\r 
\E
This current has its own superpotential $\U_{\xi} := \xi^\r \U_\r $. In 
the Palatini case the latter becomes  
the Komar superpotential \c{Ko} after suitable  modification by the frame 
change, it has the property to be a 
tensor.
The Palatini superpotential differs from the affine 
superpotential by a  contribution
induced by the choice of coordinate frame: one must compensate the 
change of cordinates by a local
linear transformation and this mixes the energy-momentum tensor and 
the $gl(D)$ current.
Finally as
explained in the previous section the antisymmetry in the two 
indices $\m$ and $\n$ of $U_\r^{\m\n}$ is spoiled by
 the presence of higher 
derivatives present in the second order formalism. The reconciliation 
of first and second order formalisms requires also some mixing with another 
symmetry in the case of the orthonormal 
frame choice, that is in the Cartan-Weyl formalism: one can check that a 
compensating Lorentz transformation allows us to relate the superpotentials 
of the two.

The whole picture can be studied  for the three theories
above Hilbert's scalar action as
was just presented or for the corresponding theories above the Einstein metric
action which is noncovariant but has only first order derivatives of 
the metric. The Einstein action 
differs from Hilbert's by a surface term and leads to the Einstein 
energy-momentum complex sometimes called pseudotensor. 
It was a big surprise when Freud \c{vF} discovered the relevant 
local superpotential,
its origin was clarified by Bergmann but it could have been conjectured
 by Noether and Hilbert! Surface terms
have been considered also in the Hamiltonian formalism \c{RT} and for 
the path integral quantization they are reviewed in \c{CN}.
We identify on the Einstein side both the Freud superpotential and the 
Sparling one in section 5.1.  In the rest of section 5 we consider the issue 
of boundary conditions and surface terms building on the previous examples. 

Let us recall that the gauge field part of the superpotential and hence 
the corresponding part of the 
current do vanish either when there is no propagating
gauge field but only a compensator or when there is a
global (bulk) spacelike Killing vector or its analog in a general gauge theory.
The asymptotic symmetry of the set of allowed configurations (and solutions), 
at the ``end" of spacetime where one does the experiment, is needed to define 
global charges because we need distinguished subgroups at infinity.
It turns out  that the contribution to these charges
from the gauge fields vanishes asymptotically despite their infinite range in 
the case of spatial Killing vectors defined also in the bulk, at least near 
infinity.

We list along the way some projects for part II.

\bigskip
\section{The general formalism and first examples}
\subsection {The general formalism}
\bigskip
A local action that depends for simplicity on the fields and their first
 derivatives \cal{S} $=\int_M L(\f,\p \f) $ 
may be invariant under a continuous (Lie) transformation. In this case one 
has:

\B \d S=0 \Leftrightarrow \d L= \p_\m S^{\m}=\frac{\p L}{\p\f}\d\f\ + \frac{\p L}{\p\p_\m\f}\d\p_\m\f
\lb{noe1} \E

\noindent Using the fact that $\p_\m$ and $\d$ commute, we obtain

\B   \p_\m S^{\m} =
[\frac{\p L}{\p\f}-\p_\m \frac{\p L}{\p\p_\m\f}]\d\f+\p_\m[\frac{\p L}{\p\p_\m\f}\d\f] \lb{noe2} \E

\noindent This implies the existence of a conserved Noether current $J^{\m}$ for each 
generator of the Lie group:

\B   J^\m := S^\m - \frac{\p L}{\p \p_\m \f} \d \f  \lb{cur1} \E

\B \p_\m J^\m \approx 0 \lb{cur11} \E

\noindent where $\approx$ means on shell. Note that $S^\m$ is not uniquely defined without more choices.

The classical theorem expresses the conservation of this current as a consequence of the
Euler-Lagrange field equations. In differential form notations $J^\m$ has a 
Hodge-dual 
$(D-1)$ form noted by $\mathbf{J}$ (where D is the spacetime dimension). 
$\mathbf{J}$ is a local function of the fields but we can only deduce from
its closedness ($\mathbf{dJ}$ $\approx 0$) that it is exact
($\mathbf{J \approx dU}$)  
if spacetime is contractible and for a given solution of the equations of motion, in particular the 
$(D-2)$ form $\mathbf{U}$ is not guaranteed to be ``local", i.e., can be 
written locally in terms of the fundamental fields of the theory.  The total charge 
$Q=\int_{V_{D-1}} \mathbf{J} $ is conserved
given sufficient decay at spatial infinity, more covariantly $\int_{\p V_D} \mathbf{J}$
$ \approx 0$ ($V_{D-1}$ is a space like hypersurface).
The addition of a topological term to the Noether current is allowed if its topological charge vanishes
so that the Noether charge is unaffected.

\bigskip

\underline{Gauge theory: the cascade equations}

Let us look at the case of a general gauge symmetry. That means that the transformation of the fields can be parametrized by a local parameter $\xi^A (x)$ (here $^A$ will denote an internal or spacetime index) and its derivatives, for example:

\B \d\f = \xi^A \Delta_A(\f) + \p_\n \xi^A \Delta^\n_A(\f)  \lb{tran1} \E

Note that this is just a special case. In fact there could be more terms with an arbitrary number of derivatives of $\xi^A (x)$.
The surface term can also be expanded in a similar way:

\B S^\m = \xi^A \Sigma_A  ^\m (\f) +  \p_\n \xi^A \Sigma_A  ^{\m\n} (\f) \lb{tran2} \E

If we insert this decomposition in \ref{cur1} and \ref{cur11} we simply 
obtain after a trivial rearrangement,

\B \p_\m (\xi^A J^\m_A  + \p_\n \xi^A U^{\m\n}_A  ) \approx 0 \lb{casbas}\E

\noindent where 

\B J_A ^\m := \Sigma ^\m_A - \frac{\p L}{\p\p_\m\f} \Delta _A(\f)  \lb{defj} \E

\B U^{\m\n}_A  := \Sigma^{\m\n}_A  - \frac{\p L}{\p\p_\m\f}\Delta^\n _A(\f) \lb{cur2} \E

Note that $J_A ^\m$ which is the coefficient of the undifferentiated $\xi^A(x)$ 
term in the total current $J^\m$ (see equations \ref{cur1} and \ref{casbas}
)  is nothing more but the usual Noether current. In fact, we recover 
the well known result by just putting $\xi^A(x) = C^t$ in 
\ref{casbas}.
The extra information due to the locality of the symmetry is 
encoded in the cascade equations which follow from \ref{casbas} by
using the arbitrariness and independence of $\xi^A(x)$ and their derivatives:

\B \p_\m \p_\n \xi^A (x) \left[ U^{\m\n}_A  \right] = 0 \Rightarrow  
U^{\m\n}_A = U^{[\m\n]}_A \lb{deb1} \E

\B \p_\m \xi^A (x) \left[ J_A ^\m +\p_\n U^{\n\m}_A \approx 0 \right] \E

\B \xi^A (x) \left[  \p_\m  J^\m_A   \approx 0 \right] \E

As usual, $(...)$ means symmetrization of the indices and $[...]$ 
antisymmetrization.
Note that the first equation is an identity whereas the other two are just on-shell equations (as emphasized by the $\approx$ symbol).

The main result of this computation is that the so-called Noether current 
$J_A ^\m$ is $\mathit{locally\ exact}$ modulo the equations of motion when the symmetry is 
local. The corresponding superpotential $U^{\n\m}_A $ has to be antisymmetric 
and can be computed directly from the Lagrangian of the theory by the use of 
equation \ref{cur2}. This antisymmetric property is particular to the case with at most first order derivatives of the fields.

\bigskip

\underline{The Noether identities}

We would like here to recall the famous second Noether theorem which gives 
some relations between equations of motion when some gauge symmetry is present. 
We will show how they can be deduced in our formalism. If we use the  decompositions \ref{tran1} and \ref{tran2}, the definitions \ref{defj} and \ref{cur2} 
in the equation \ref{noe2} we obtain that 

\B \frac{\d L}{\d \f} \left( \xi^A \Delta_A + \p_\n \xi^A \Delta^\n_A  \right) = \p_\m \left( \xi^A J^\m_A  + \p_\n \xi^A U^{\m\n}_A  \right) \lb{idno} \E

where $\frac{\d L}{\d \f}$ are just the Euler-Lagrange equations derived from 
$\f$ and an abstract summation over all the fields of the theory is understood. These equations are now exact and we can look for the cascade equations corresponding to this equality:

\B \xi^A \hspace{.3in} \left[ \frac{\d L}{\d \f} \Delta_A =  \p_\m J^\m_A \right] \lb{idcas1} \E

\B \p_\m \xi^A \hspace{.3in} \left[  \frac{\d L}{\d \f} \Delta^\m_A = J^\m_A + \p_\n U^{\n\m}_A \right] \lb{idcas2} \E

\B \p_\m \p_\n \xi^A \hspace{.3in} \left[ U^{\m\n}_A = 0 \right] \lb{idcas3} \E

These last equations replace the cascade equations when no use of the equations of motion is permited. We now see why equation \ref{deb1} was exact. 
If we now replace $J^\m_A$ as given by equation \ref{idcas2} into \ref{idcas1} and make use of the antisymmetry of $U^{\m\n}_A$ we easily obtain the Noether identities:

\B \frac{\d L}{\d \f} \Delta_A(\f) = \p_\m \left( \frac{\d L}{\d \f} \Delta^\m_A(\f) \right) \lb{2no} \E

Note that the Noether identities do not depend on any surface term because all of 
them are hidden in $J^\m_A$.
This old result will be used in section 4 for a better understanding of the 
affine gauge theory and its reduction to Einstein theory.

Remember that we have just treated the simple case where the decompositions 
\ref{tran1} and \ref{tran2} go only up to first derivative in $\xi^A(x)$. In 
a more general case, the above conclusions have to be modified as we will see in 
section 3 in the specific example of General Relativity. 

\bigskip

\underline{The abelian cascade: the $U^{\m\n}_\xi$ superpotential}

In the next subsection we will give some examples but before that, let us just 
show that $J^\m (\xi^A (x))$ of \ref{cur1} (noted now simply $J^\m_\xi$) where $\d \f$ is given by \ref{tran1} and $\S^\m$ by \ref{tran2}.
can be expressed as a divergence. In fact the corresponding (local)parameter 
dependent superpotentials $U^{\m\n} (\xi^A (x))$ (noted $U^{\m\n}_\xi$) will be 
the most important object in gauge theories to compute conserved charges as we 
shall see in specific examples. Let us use the decomposition

$$\xi^A (x) := \e (x)  \xi_0^A(x)  $$

in equation \ref{casbas}. $\e (x)$ is just the local parameter for an 
 abelian \footnote{ Actually the subgroup is not really abelian in the case of 
diffeomorphisms but there all changes of coordinates are linearly related or 
unidimensional.}
 subgroup with $\xi_0^A(x)$ fixed.  The cascade 
in terms of $\e (x)$ and its derivatives gives after some trivial algebra the main result:

\B  J^\m_{\xi_0} \approx - \p_\n  U^{\n\m}_{\xi_0}  \E

where in this case $U^{\n\m}_{\xi_0}$ is simply $U^{\n\m} _A  \xi_0^A$. As we shall see the case of General Relativity in its 2nd order formulation (section 3) is only slightly more 
difficult but the idea is similar. We will introduce its affine formulation in 
section 4 where the computations will be easier and the comprehension more 
profound maybe.
Let us insist that $U^{\n\m}_{\xi_0}$ is the fundamental object we shall use to 
compute physical charges in gauge theories. The main difficulty is to select 
the appropriate ${\xi_0}$'s to get gauge invariant results. 

\bigskip

\underline{The conserved charges}

The usual Noether conservation law $\ex \J \approx 0$ can be used to define a 
conserved charge. For that purpose
 we may integrate this equation  on a D-dimensional 
spacetime bounded by two spatial hypersurfaces (say $\Si_1$ and 
$\Si_2$) and by spatial infinity (say $\Si_\infty$). If we impose the physical 
condition that $\J$ has a vanishing flux through
 $\Si_\infty$ (in other words that 
the charge does not leave spacetime between $\Si_1$ and $\Si_2$) , Stokes' theorem implies that 
$\int_{\Si_1} \J = \int_{\Si_2} \J$ and so the charge defined as the integral of
 $\J$ on a spatial hypersurface is conserved and is independent on the choice of spacelike hypersurface. 

The case of a local symmetry is more subtle because the usual Noether 
conservation may be replaced by 

\B \J \approx \ex \U \lb{eqb} \E

This has radical consequences for the meaning of what is a conserved quantity 
and how to define it. In fact now equation
 \ref{eqb} can be integrated on a 
$(D-1)$ dimensional manifold, in two different ways:

\VS

- \underline{On $\Si_\infty$} : This will be the right choice to define a 
conserved charge in General Relativity and Yang-Mills theories. Let $B_{1\infty}$ and $B_{2\infty}$ the boundaries of 
$\Si_\infty$ at time $t_1$ and $t_2$ respectively (these are actually D-2 
dimensional closed manifolds). If we again assume that the flux of $\J$ vanishes on 
$\Si_\infty$ then Stokes law applied to equation \ref{eqb} will imply that 
$\int_{B_{1\infty}} \U = \int_{B_{2\infty}} \U$. Then the conserved charge may 
be defined as the integral of $\U$ on the infinite spatial boundary of a 
time-fixed hypersurface:

\B Q = \int_{B_{\infty}} \U \lb{loccha} \E

This definition is completely independent of the fact 
that there exist or not an 
interior black hole horizon or singularity
 inside space time. The key point is that this construction never leaves the
asymptotic region and is both robust and physical as that is precisely where charges are measured. As we will discuss in more 
detail in the next examples (Yang-Mills and Gravitation), there exist relations
 between the boundary conditions we have to impose on our fields to define the variational principle,  the form of
asymptotic Killing vectors and the associated gauge invariant conserved charges.
 In some very special 
cases (for instance in  presence of a global spatial Killing vector), we should 
be able to use other timelike hypersurface than $\Si_\infty$, for instance at 
finite distance (say $\Si_{r_0}$), leading to the notion of 
quasi-local charges. We postpone this discussion to section 5.2 for the gravitational case.

\VS

- \underline{On $\Si_1$} : In that case, we will obtain relations between 
quantities computed at a fixed time. Take for example the next simplest case 
where spacetime has one interior boundary $B_{1H}$ (for instance a black hole
horizon). Then we will obtain the general relation:

\B \int_{B_{1\infty}} \U = \int_{B_{1H}} \U + \int_{\Si_1} \J \lb{posthe} \E

This kind of equation has been used successfully in General Relativity, 
for instance to prove the positivity of the ADM mass and the first law of 
thermodynamics. Our purpose here is not to repeat nor give other demonstrations 
of 
these crucial results. We will just show in section 5.2 that their starting point 
is nothing but
 a well understood version of equation \ref{posthe} which is a direct 
consequence of the locality of diffeomorphism invariance. 

Let us anticipate
and remark here that the left hand side is the charge defined by the assumption of asymptotic symmetry of the fields, which allows us to avoid the nogo theorem of Hilbert and Noether. However there is no general prescription yet to define 
separately either of the terms on the right hand side, this will be studied in 
paper II of this series.
Clearly the choice of inner boundary has to obey the zero flux condition and 
imust involve some dynamical knowledge. This could be useful for the present 
sudies of anti de Sitter spaces.  

\bigskip

\subsection{Yang-Mills case}

\bigskip

\underline{General formalism}:

\VS

Let us start with the usual Yang-Mills Lagrangian, eventually coupled to a matter term:

\B L_{YM} = - \frac{1}{4} F^{\m\n}_A F_{\m\n}^A + L_{mat}(\f,\p_\m \f, A_\m^A) \lb{ymlag} \E

where $A_\m^A $ is the gauge potential, $F_{\m\n}^A$ its associated curvature 
and $\f$ a matter field lying in some representation $R$ 
of the gauge group. We also assume that $L_{mat}$ depends on $A_\m^A$ only through the covariant derivative of $\f$.

This Lagrangian is invariant under the local gauge transformation:

$\d_\xi A_\m^A = \p_\m \xi^A + f^A_{\ BC} A_\m^B \xi^C = D_\m \xi^A$

$\d_\xi \f = \xi^A  R_A \f $

$ \Rightarrow \d_\xi L_{YM} = 0 $

Of course $f^A_{\ BC}$ are the structure constants and $R_A$ the specific 
infinitesimal generators in the representation $R$ of the group.

We can now use this symmetry in equation \ref{cur1} and then rewrite it as in 
equation \ref{casbas}. The useful quantities \ref{defj} and 
\ref{cur2} can now be computed for the Yang-Mills Lagrangian \ref{ymlag} :

$J^\m_A =- f^C_{\ BA} A_\n^B F^{\m\n}_C + \frac{\p L_{mat}}{\p\p_\m\f} R_A \f $

$ U^{\m\n}_A =  F^{\m\n}_A $

$J^\m_\xi = \xi^A J^\m_A + \p_\n \xi^A U^{\m\n}_A$

We see that $F^{\m\n}_A$ is just the superpotential of the naive Noether current $J^\m_A$.
Now the cascade equations are nothing more but the Yang-Mills equations:

$\p_\n (F^{\m\n}_A) \approx -f^C_{\ BA} A_\n^B F^{\m\n}_C + \frac{\p L_{mat}}{\p\p_\m\f} R_A \f$

Our purpose is to study the fate of conserved quantities in the presence of
 local invariance so it is important to recognize these equations as 
 conservation equations of the type $\J \approx \ex \U$.
Here we have a superpotential which is not anymore a gauge scalar. In fact the 
integral at spatial infinity of $F^{\m\n}_A$ does not make any sense as a 
conserved quantity because it is not gauge invariant. The good gauge independent
superpotentials are thus parameter dependent ones and can be obtained by the 
abelian cascade method. The result is obvious:

\VS

$U^{\m\n}_\xi = F^{\m\n}_A \xi^A $

$J^\m_\xi \approx \p_\n U^{\m\n}_\xi$

\VS

If we recall the discussion of conserved charges of the previous subsection, we obtain that the gauge invariant conserved charge is (equation \ref{loccha}) :

\B Q(\xi^A(x)) = \int_{B_\infty}  \U_\xi  \lb{ymcha}\E

where as usual $\U_\xi$ is the $\mathit{D-2}$ form associated to the Hodge dual of $U^{\m\n}_\xi$.

The point is now that in order to obtain physical charges we have to specify 
and select what $\xi^A (x)$ can be. This is treated in the following.

\bigskip

\underline{The Yang-Mills Charges}

First we would like to recall an important point which has to be taken into 
account in a variational principle. A variational principle is defined only when 
boundary conditions are specified. In addition, if a boundary condition is chosen, we 
cannot add anymore an arbitrary total derivative to the Lagrangian because in general 
when the fundamental fields of the theory do not vanish on the boundary (say at 
infinity) the variational principle (i.e. $\d S = 0 \Rightarrow$ Equations of motion) 
will not be satisfied.

For example, the variational principle for the Yang-Mills Lagrangian \ref{ymlag} implies that 

\B \int_{\p M} \frac{\p L}{\p\p_\m A_\n^A} \d  A_\n^A +\frac{\p L}{\p\p_\m\f} \d \f \lb{varcon}\E

has to vanish for an arbitrary variation. We do not want to analyse here the behaviour of the solutions of this equation say at spatial infinity in terms of power series in $\frac{1}{r}$. We will consider the simplified Dirichlet case as if infinity was at a finite distance like in a compactification of spacelike infinity:

\[ \lim_{r \rightarrow \infty} \d A_\n^A = 0 \]

\[ \lim_{r \rightarrow \infty} \d \f = 0 \]

The full mathematical analysis is deferred to our second paper.

If we use this for the special case of a gauge variation, we obtain the boundary ``Killing" equations:

\B \lim_{r \rightarrow \infty} D_\m \xi^A =0 \lb{diri1} \E

\B \lim_{r \rightarrow \infty} \xi^A  R_A \f = 0 \lb{diri2} \E

The last two equations tell us which asymptotic $\xi$'s are allowed in equation \ref{ymcha}. These  can form an infinite asymptotic group (see for instance the gravitational case) but only their asymptotic form is used and these can be a finite number of those charge by means of equation \ref{ymcha}. 

Note also that equation \ref{varcon} used for the simple case $\d =\d_\xi$ will imply the vanishing of $J^\m_\xi$ at spatial infinity and so the existence of a conserved charge showing the consistency of our framework. We can go even further in the analysis when there exists some global Killing parameter (i.e. $D_\m \xi_K^A =0$). In that case  $J^\m_{\xi_K} = 0$ everywhere and so the corresponding charge $Q(\xi_K)=\int_B U^{\m\n}_{\xi_K}$ can be computed on any (D-2) dimensional surface outside matter sources.

We want to insist here on the following points

- The case of Dirichlet conditions \ref{diri1} and \ref{diri2} is just the simplest 
solution for the vanishing of equation \ref{varcon}. The general solution to this 
condition has to be treated in the asymptotic regime with the appropriate decrease.

- Physical conditions will specify the boundary condition (as in the case of free or fixed-ends strings) which will not only fix part of the surface term of the Lagrangian but also give some conditions on the asymptotically allowed gauge parameters.

-Boundary conditions should be gauge invariant. In the case of General Relativity this is made possible by introducing a reference space at infinity (where it is needed).

\VS

Some well known examples are:

- The Maxwell case with matter fields which vanish at infinity. In that case the asymptotic Killing equation just becomes $\lim_{r \rightarrow \infty} \p_\m
\xi = 0$. The subalgebra which will give a non vanishing finite charge will be $\Re$. Thus the number of charges is just 1 (the dimension of $\Re$, which is also the number of independent Casimir operators of the subgroup). In addition, $\xi = C^t$ is a global Killing parameter and so we recover the well known result that the electric charge can be computed on any closed surface which surrounds the charged matter distribution.

- The SU(2) Yang-Mills-Higgs system where a particular solution to the asymptotic Killing equations is just $\xi^A = \F_0^A$ (the direction of the Higgs field at infinity), see for instance \c{AD}.

\subsection{The p-form theory}

We can consider the abelian p-form Lagrangian given essentially by

\B \L = \Gb \we \st \Gb \E

Where $\Gb = \ex \Bb$, $\Bb$ being the p-form abelian gauge field (see \c{Ju}).

The local gauge invariance is just $\d_{\xib} \Bb = \ex \xib$, where $\xib$ is an arbitrary (p-1)-form gauge parameter. We will not repeat all the computations but just give the final result which is that the parameter dependent conserved charge is given by:

\VS

$Q(\xib)=\int_{B_\infty} \xib \we \st \Gb $

\VS

If we again impose Dirichlet type boundary conditions, the analogue of \ref{diri1} and \ref{diri2} for $\xib$ is thus:

\[ \lim_{r \rightarrow \infty} \ex \xib = 0 \]

It is obvious from the definition of $ Q(\xib)$ and the equations of motion of $\Bb$ ( $\ex \st \Gb \approx 0$) that when $\xib = \ex \bb$, the charge will vanish on shell (remember that $B_\infty$ is already a boundary hence is closed and that partial integration can be done without any boundary term). 
Thus in the case of the p-form, the subgroup which could potentially give some non trivial conserved charge is just the set of (p-1)-forms which are closed but not exact or in other words the $\mathit{(p-1)}^{th}$ De Rham cohomology group $H^{p-1}$ of $B_\infty$. The number of conserved charges will then be given by the $\mathit{(p-1)}^{th}$-Betti number $b^{p-1}(B_\infty)=dim \left( H^{p-1} (B_\infty) \right)$.  For example for a spacetime with 2 infinite boundaries components (wormhole) we recover 2 ordinary charges.

\VS

The reader interested only in fluid dynamics can now skip to section 6.

\section{The classical case of General Relativity}

\bigskip

\subsection{Second order form of gravitation: the cascade Equations for diffeomorphisms}

\bigskip

Let $L(g, \p g, \p^2g) = \frac{1}{2k} \sqrt{-g} R$ be the scalar Hilbert Lagrangian density of our theory. It is equal to the so-called Einstein Lagrangian up to the surface term that eliminates second derivatives of the metric, see section 5. A variation of L is given by

\B  \d L = \frac{\p L}{\p g} \d g + \frac{\p L}{\p \p_\m g} \d \p_\m g +\frac{\p L}{\p \p_\m \p_\n g} \d \p_\m \p_\n g \lb{var2}\E

Where we omitted the spin indices of $g_{\a \b}$ for notational simplicity. Using the second order equations of motion $\frac{\d L}{\d g} =\frac{\p L}{\p g} -\p_\m \frac{\p L}{\p \p_\m g} +\p_\n \p_\m \frac{\p L}{\p \p_\n \p_\m g} \approx 0$
we can write equation \ref{var2} as a total derivative,

\B  \p_\m J_\xi^\m := \d L - \p_\m \left(\frac{\p L}{\p \p_\m g} \d g - \p_\n \left( \frac{\p L}{\p \p_\m \p_\n g} \right) \d g + \frac{\p L}{\p \p_\m \p_\n g} \p_\n \d g  \right) \approx 0\lb{tder} \E

Our Lagrangian density is again such that the action is invariant under a reparameterization $x^\r \rightarrow x^\r +\xi^\r (x)$. Putting the well known expressions for the variation $\d L = \li_\xi L = \p_\r(\xi^\r L)$ and $\d g_{\a\b} = \li_\xi g_{\a\b} = \xi^\r \p_\r g_{\a\b} + \p_\a \xi^\r g_{\r\b} + \p_\b \xi^\r g_{\a\r}$ in \ref{tder} and sorting out the factors of $\xi^\r$, $\p_\n \xi^\r$ and $\p_\d \p_\n \xi^\r$, we obtain

\B  \Leftrightarrow \p_\m (\xi^\r T_\r ^{\ \m} + \p_\n \xi^\r U_\r^{\ \m\n} +  \p_\d \p_\n \xi^\r V_\r^{\ \m (\n \d)}) \approx 0 \lb{eqbase} \E

Where $T_\r^{\ \m}$ and  $U_\r^{\ \m \n}$ are the canonical energy-momentum complex (called sometimes canonical energy-momentum pseudotensor and noted $t_\r^{\ \m}$) and the canonical spin complex respectively,

\VS

\VS

$ T_\r^{\ \m} \hspace{.20in}   := \d _\r^\m L - \left(\frac{\p L}{\p \p_\m g_{\a \b}}-\p_\n \frac{\p L}{\p \p_\m \p_\n g_{\a \b}}\right) \p_\r g_{\a \b} - \frac{\p L}{\p \p_\m \p_\n g_{\a \b}}
\p_\n \p_\r g_{\a \b} $

$\hspace{.41in}    =\frac{\sqrt{-g}}{2k} \left( R \d_\r^{\ \m}+\G^\a_{\a\b,\r}g^{\b\m}-\G^\m_{\a\b,\r}g^{\a\b} \right)$

\VS

$  U_\r^{\ \m \n} \hspace{.05in} := \left( \p_\eta \frac{\p L}{\p \p_\m \p_\eta g_{\a \b}} - \frac{\p L}{\p \p_\m g_{\a \b}} \right) \Lambda_{\a \b}^{\n \g} g_{\r \g} - 
\frac{\p L}{\p \p_\m \p_\n g_{\a \b}} \p_\r g_{\a \b} - \frac{\p L}{\p \p_\m \p_\eta g_{\a \b}} \Lambda_{\a \b}^{\n \g} \p_\eta g_{\r \g} $

$\hspace{.41in}     =\frac{\sqrt{-g}}{2k} \left[  \d_\r^{\ \m} \G^\n_{\a\b} g^{\a\b} + \G^\a_{\a\r}g^{\m\n} - 2 \G^\m_{\a\r} g^{\n\a} \right]$

\VS

$ V_\r^{\ \m (\n \d)} := - \frac{\p L}{\p \p_\m \p_\d g_{\a \b}} \Lambda_{\a \b}^{\n \g} g_{\r \g} $ (symmetrized in $^{\n\d}$)

$\hspace{.41in}     =\frac{\sqrt{-g}}{2k}  \left[\frac{1}{2} g^{\m\d} \d^{\ \n}_\r + \frac{1}{2} g^{\m\n} \d^{\ \d}_\r - g^{\d\n} \d^{\ \m}_\r \right]$

\VS

$\Lambda_{\a \b}^{\n \g} := \d_\a^\n\  \d_\b^\g + \d_\a^\g\ \d_\b^\n$ and $\G = \G^{(g)}$ is the Levi-Civita connection.

Then, as in the previous examples, we derive the cascade Equations:

\B  \p_\m \p_\n \p_\d \xi^\r  \left[ V_\r^{\ \m \n \d} \right] = 0 \Leftrightarrow V_\r^{(\m \n \d)} = 0 \lb{2cas1} \E

\B  \p_\m \p_\n \xi^\r  \left[  U_\r^{\ \m \n} + \p_\d V_\r^{\ \d \n \m} \right] = 0  \Leftrightarrow U_\r^{\ \m \n} + \p_\d V_\r^{\ \d \n \m} = F_\r^{[\m \n]}\lb{2cas2} \E

\B  \p_\m \xi^\r  \left[ T_\r^{\ \m} + \p_\n U_\r^{\ \n \m} \right] \approx 0\lb{2cas3} \E

\B  \xi^\r \left[ \p_\m T_\r^{\ \m} \right] \approx 0 \lb{2cas4} \E

And 

\B F_\r^{\m \n} := \frac{\sqrt{-g}}{2k} \left[ 2 \d^{\ \m}_\r \G^\n_{\a\b} g^{\a\b} -2 \d^{\ \n}_\r \G^\m_{\a\b} g^{\a\b} +g^{\m\a}\ \G^\n_{\r\a}  -g^{\n\a}\ \G^\m_{\r\a}  \right] \lb{Vsurf} \E

Note that $U_\r^{\ \n \m}$ is not antisymmetric and that the first two equations are exact but that the last two are on-shell.

We again see an important fact: equation \ref{2cas3} shows that the current (in this case the canonical energy-momentum complex) can be written down as a divergence. The gauge diffeomorphism invariance of General Relativity implies then that the charge associated with this symmetry may be expressed as a surface integral.

We already saw in the case of Yang-Mills theory that the most important quantity to define conserved charges is the (local) parameter-dependent superpotential. To derive it, we will use the abelian cascade trick in this non trivial example. We shall just give the formal result without trying to obtain physical consequences for the moment. We shall postpone this question to section 5.2. The motivation to construct such an object is just that it will provide a single formula for conserved quantities like total mass or angular momentum. 
The point is that none of the tricks Landau-Lifshitz or Weinberg used to construct  the gravitational angular momentum starting from a symmetrized canonical energy-momentum complex is needed. The connection with the Komar or Katz superpotentials will be established in the following sections using the affine gauge formalism, where it is much simpler.

\VS

Let us start with equation \ref{eqbase} :

\B \p_\m J^\m _\xi \approx 0 \lb{epcas}\E

where $J^\m _\xi = \xi^\r T_\r ^{\ \m}  + \p_\n \xi^\r U_\r^{\ \m\n} +  \p_\d \p_\n \xi^\r V_\r^{\ \m \n \d} $

Now, let us define 
$\xi_0^\r (x) \e (x) = \xi^\r (x)$. Again, $\e (x)$is a local parameter for an abelian subgroup. $\xi_0^\r (x)$ is kept fixed and will be  determined for each conserved charge to be computed. 
Using this decomposition in equation \ref{epcas} we get that

\B \p_\m ( \e\hspace{.01in} J^\m _{\xi_0} + \p_\n \e\hspace{.01in} U^{\m\n}_{\xi_0} +\p_\d \p_\n \e\hspace{.01in} V^{\m\n\d}_{\xi_0} ) \approx 0 \E

The abelian cascade equations are computed in terms of derivatives of $\e$ and give:

\B \p_\m \p_\n \p_\d \e: \hspace{.02in}  V_{\xi_0}^{(\m \n \d)} = 0 \lb{e2cas1} \E

\B  \p_\m \p_\n \e: \hspace{.02in}   U_{\xi_0}^{\ \m \n} + \p_\d V_{\xi_0}^{\ \d \n \m} = F_{\xi_0}^{[\m \n]}\lb{e2cas2} \E

\B \p_\m \e : \hspace{.02in}    J_{\xi_0}^{\ \m} + \p_\n U_{\xi_0}^{\ \n\m} \approx 0\lb{e2cas3} \E

\B \e: \hspace{.02in}    \p_\m J_{\xi_0}^{\ \m}  \approx 0\lb{e2cas4} \E

So equation \ref{e2cas3} shows, as one expects, that the total Noether current $J^{\ \m}_{\xi_0}$ can be written in terms of a divergence modulo equations of motion due to the locality of the symmetry. For completeness let us write down the formula for $U^{\m\n}_{\xi_0}$ in the gravitational $2^{nd}$ order formalism:

\B U_{\xi_0}^{\ \n \m} = \xi_0^\r U^{\n\m}_\r +2 \p_\d \xi_0 ^\r  V_\r ^{\n \m \d } \lb{esup} \E

In what follows, we will omit the $_0$ subscript from $\xi_0$.  

\VS

We may conclude this discussion by just giving the connection of the above formulas with some well known results:

-  $T_\r ^{\ \m}$ is nothing but one half the originally rescaled M\o ller 
 energy-momentum pseudotensor \c{Mo}.

- This pseudotensor can be written as the divergence of the canonical spin complex which is not antisymmetric (equation \ref{2cas3}). However there exists an antisymmetric superpotential which does the same job:

\B  \ _MU^{\ \m\n}_\r  := U^{\ \m\n}_\r - \p_\d W^{\ \n[\m\d]}_\r \lb{luck} \E

\B W^{\ \n[\m\d]}_\r := \frac{\sqrt{-g}}{2k} \left( g^{\n\m}\d^{\ \d}_\r-g^{\n\d}\d^{\ \m}_\r \right) \lb{wdef} \E

hence equation \ref{2cas3} and the above definitions imply that

\B T_\r^{\ \m} \approx \p_\n \ _MU_\r^{\ \m \n} \E

Where $\ _MU^{\ \m\n}_\r$ is one half the superpotential introduced (and rescaled) by M\o ller \c{Mo}, and is equal to:

\B  _MU_\r^{\m\n} = \frac{1}{2k} \sqrt{-g}g^{\m\a}g^{\n\b} ( \p_\a g_{\b\r} - \p_\b g_{\a\r} ) \lb{mol2} \E

$$ \hspace{.35in} = - \frac{1}{2k} \sqrt{-g} \left( \G^\m_{\a\r}\ g^{\a\n} - \G^\n_{\a\r}\ g^{\a\m} \right) $$

The same is true for $U^{\m\n}_\xi$, equation \ref{esup}: there exists an antisymmetric version of it,

\B  \ _KU^{\ \m\n}_\xi := U^{\ \m\n}_\xi - \p_\d \left( \xi^\r W^{\ \n[\m\d]}_\r \right) \lb{luck2} \E

\B J_\xi^{\ \m} \approx \p_\n \ _KU^{\ \m\n}_\xi \E

Where now $\ _KU^{\ \m\n}_\xi$ is one half the Komar \c{Ko} superpotential,

\B  _KU_\xi^{\m\n} := \frac{1}{2k} \sqrt{-g} \left( \bigtriangledown ^\m \xi^\n - \bigtriangledown ^\n \xi^\m \right) \lb{kom} \E

\B \hspace{.35in} = \ _MU^{\ \m\n}_\r \xi^\r + W^{\ \d[\m\n]}_\r \p_\d \xi^\r \lb{komp} \E

Where we used in the last equation that $V_\r^{\ \m \n \d} =\frac{1}{2} ( W _\r^{\ \n \m \d} + W _\r^{\ \d \m \n} )$.

One could think that we have been lucky that such superpotentials exist for equations \ref{luck} and \ref{luck2}. We shall show, in the Affine Gauge formalism (section 3) that their existence is due to some local symmetry, just as the existence of the canonical spin complex $U^{\ \m\n}_\r$. We will also understand in an elegant way the lack of antisymmetry of the latter. 
The point is that the Affine Gauge formalism is a first order formulation so the antisymmetry is guaranted from the begining as we saw in section 2.1. Equation \ref{komp} will then be derived in a natural way. We will also understand in an elegant way its non-antisymmetry.
Before that let us show that the addition of a matter field will not essentially change the above formulas.

\subsection{Matter's contribution: The symmetric tensor}

All the previous discussion was for vacuum gravitational theory. However we would like to add some matter, i.e. $L_{Matter} (\F , \p\F )$, and see how this can affect our equations. 

The basic Noether theorem gives a formula which allows us to calculate a conserved current coming from a global symmetry of our Lagrangian. When this symmetry is the translation invariance in a flat background then the conserved current is just the so-called canonical energy-momentum matter tensor $_ct_\r^{\ \m}$ (lower case letters will be used for the contribution coming from the matter fields), which is given by the usual formula:

\B  _ct_\r^{\ \m} = \d_\r^{\ \m} L_M- \frac{\p L_M}{\p \p_\m \F} \p_\r \F  \hspace{0.2in}\mbox{and} \hspace{0.2in} \p_\m\ _ct_\r^{\ \m} \approx 0 \E

Then, the time conserved physical quantity is the 4-momentum vector $P_\r = \int_V\ _ct_\r^{\ 0} dV$.

In general, this energy-momentum tensor is not symmetric. As Belinfante has shown \c{Be} , it is possible to add an antisymmetric surface term to symmetrize it without changing the physics, i.e. 

\B  _Bt_\r^{\ \m} =\ _ct_\r^{\ \m} + \p_\n\ _B\Sigma^{\ [\m\n]}_\r  \E

and $\int_\infty \Sigma^{\ [0 i]}_\r dS_i \longrightarrow 0$, $ _Bt^{\sigma\m} =\eta^{\sigma\r}\ _Bt_\r^{\ \m} =\ _Bt^{\m\sigma}$.

\VS

When our background space-time becomes a dynamical variable $g_{\a\b}$, then the matter energy momentum  tensor $_st^{\sigma\m} = 2 \frac{\p L_M}{\p g_{\sigma\m}}$ appears as the source in Einstein's equations. The symmetry of its upper indices is guaranteed by the symmetry of the metric and it has been verified that this quantity effectively coincides with the above Belinfante version. 
In what follows, we shall give another proof of this important fact and see how this can affect the total energy-momentum of the gravitational part. The antisymmetric surface term needed to symmetrize $_ct_\r^{\ \m}$ appears naturally as the $\mathit{matter\ superpotential}$.

Let $L_M (g,\F_m,\p \F_m)$, the scalar Lagrangian density, be a functional of a set of fields $\F_m$ (where $_m$ is a spin index) and their first derivatives and of the background metric. We now use the fact that we can write the matter variation of the Lagrangian in a total derivative form by making use of the equations of motion of $\F_m$ only:

\B \d L_M \approx \p_\m \left( \frac{ \p L_M}{\p \p_\m \F_m} \d\F_m\right) + \frac{ \p L_M}{\p g_{\sigma\m}} \d g_{\sigma\m} \lb{matiere} \E 

The variations of all the fields are given by their Lie derivative. Let us define the matrix $\Delta$ which acts on the spin index of $\F_m$ by:

\B \d\F_m = \xi^\r \p_\r\F_m + \p_\n \xi^\r (\Delta^{\ n}_m )_\r^{\ \n} \F_n \lb{soix} \E

\VS

Equation \ref{matiere} becomes

\B  \p_\m \left( \xi^\r\ _ct_\r^{\ \m} + \p_\n \xi^\r u_\r^{\ \n\m} \right)  \approx \frac{_st^{\sigma\m}}{2} (\xi^\r \p_\r g_{\sigma\m} + 2 \p_\m \xi^\r g_{\sigma\r}) \E

Where $u_\r^{\ \n\m} := - \frac{ \p L_M}{\p \p_\m \F_m} (\Delta^{\ n}_m )_\r^{\ \n} \F_n$ is the $\mathit{matter\ superpotential}$ and $_ct_\r^{\ \m}$ and $_st^{\sigma\m}$ have been defined above.
Now, the cascade equations are 

\B  \p_\n \p_\m \xi^\r \left[ u_\r^{\ \n\m} \right] = 0 \Leftrightarrow u_\r^{\ \n\m}=u_\r^{[\n\m]} \lb{mcas1}\E

\B  \p_\m \xi^\r \left[_ct_\r^{\ \m} + \p_\n u_\r^{\ \n\m} \approx \ _st^{\sigma\m} g_{\sigma\r} \right] \lb{mcas2}\E

\B  \xi^\r \left[ \p_\m\ _ct_\r^{\ \m} \approx \frac{_st^{\sigma\m}}{2} \p_\r g_{\sigma\m} \right] \lb{mcas3} \E

Then equation \ref{mcas2} shows that if the Euler-Lagrange equations of  $\F_m$ hold then the matter superpotential is the antisymmetric quantity that we have to add to the canonical energy-momentum tensor to obtain the symmetric one. The last equation cannot be identified with a conservation law unless $_st^{\sigma\m} \approx 0$ which is the case only when no dynamical term for the metric are present. 

Let us summarize: The addition of a matter Lagrangian to the gravitational one affects the equations of section 3.1 by just adding a symmetric tensor term to the gravitational canonical energy-momentum complex. In fact, it is easy to see that with the simple change $T_\r^{\ \m} \rightarrow T_\r^{\ \m} + t_\r^{\ \m}$, all the above equations remain unchanged, always keeping in mind that the previous vacuum equations of motion are modified by matter terms. Finally we have gained a deeper understanding of the relation betwen the canonical matter tensor and the symmetric one.
In what follows, we will return to the vacuum case. We just need to keep in mind the fact that the addition of an integer spin matter field doesn't change anything if we proceed with the above substitution. The case where the matter field is a finite spinorial representation of the Lorentz group is more subtle. 
In that case we will not be able to use our affine gauge theory (because the universal covering group of $GL(D,\Re)$ requires infinite spinorial 
representations, see for instance \c{HeM}), for the same reason that we cannot use 
Einstein formalism to deal with spinors. However, the results derived for the affine 
formalism can be reduced to the orthogonal case, allowing the addition of for example 
Dirac spinors. Let us now turn to our new (affine) first order formalism, which 
generalizes Cartan-Weyl or Palatini formalism. Note that if we deal with spinors or 
gravitinos, first order formalism can introduce torsion. So we will allow torsion and 
 even nonmetricity.

\bigskip

\section{Superpotentials of Affine gauge relativity}

\subsection{Definition of the theory}

\bigskip
The simplest mathematical form of general relativity uses moving linear frames, it is the so-called first order formalism. The classical case is the $so(1,D-1;\Re)$ formulation.
Let L(M) be the linear frame bundle over a D-dimensional Riemannian manifold M with metric $g$ ($D > 2$).
Let us consider on L(M) a Lagrangian D-form $\L$, function of a linear 1-form connection $\w^a_{\ b}$ (Yang-Mills $gl(D,\Re)$), of the canonical 1-form $\th^{a}$ ($\Re^D$ valued) and of the metric $g^{ab}$ (which will be used to lift and lower the $\Re^D$-valued indices), as well as their first derivatives:

\B  \L=\frac{1}{2k} \R^a_{\ e}  \we \sqrt{-g} g^{eb} \S_{ab}  \lb{actgrav} \E

\noindent $k=8 \pi G$ is the usual normalisation factor.
we defined

\VS

$\S_{a_1...a_r} := \frac{1}{(D-r)!} \epsilon_{a_1...a_rc_{r+1}...c_D} \th^{c_{r+1}}  \we...  \we \th^{c_D}$

\VS
\noindent where $\epsilon_{a_1...a_D}$ is the Levi-Civita tensor density symbol.
The curvature 2-form $\R^a_{\ b}$ is as usual defined by

\B  \R^a_{\ b} = \ex \w^a_{\ b} + \w^a_{\ c} \we \w^c_{\ b} \E

We can also define the torsion $\X^a$ and the nonmetricity $\M^{ab}$ as the covariant derivatives (we will say curvatures) of the moving frame and the metric respectively: 

\B  \X^a := \cex \th^a = \ex\th^a+\w^a_{\ b} \we \th^b   \E

\B \M^{ab} := \cex g^{ab} = \ex g^{ab} + \w^{ab} +\w^{ba} \E
\noindent where $\cex$ is the $gl(D,\Re)$ covariant derivative.  
We remember that the torsion measures some covariant anholonomy and the nonmetricity the failure of the metric $g^{ab}$ to be compatible with the connection. Their vanishing defines uniquely the Levi-Civita connection as a function  of the metric and its derivatives.

Due to their definitions these three ``curvatures" obey the following Bianchi identities:

\B \cex \R^a_{\ b} = 0 \lb{bian1} \E

\B \cex \X^a = \R^a_{\ b} \we \th^b \lb{bian2} \E

\B \cex \M^{ab} = \R^{ab} + \R^{ba} \lb{bian3} \E

In the previous sections we learned that to obtain some cascade equations coming from a gauge symmetry of the Lagrangian we need a variational principle. We will then take the pullback along an $\mathit{arbitrary}$ local section of the above quantities. We will keep the same letter to note the pullback quantity for notational simplification. 
The two usual sections are 

$\bullet$) the coordinate or holonomic section where $s^* (\th^a) = \ex x^a$ and 
$s^* (\w^a_{\ b} ) = \G^a_{cb} \ex x^c$ ( $\G^a_{cb}$ is an affine connection on $M$) 
and 

$\bullet \bullet$) the rigid or orthogonal section where $s^* (\th^a) = \eg^a$ and $s^* (\w^a_{\ b} ) = \g^a_{cb} \eg^c$ ($\eg^a$ are the well known orthogonal Lorentz frames, sometimes called tetrads or vielbein, and $\g^a_{cb}$ the Ricci ``rotation" coefficients).

We will allow the fields $\w^a_{\ b}$, $\th^a$ ($s^* (\th^a)$ in fact) and $g^{ab}$ to vary independently. The Euler-Lagrange equations corresponding to the Lagrangian \ref{actgrav} are

\B \frac{2k}{\sqrt{-g}}\  2 \frac{\d \L}{\d g^{ab}} = \R^c_{\ a} \we \S_{cb} + \R^c_{\ b} \we \S_{ca}- g_{ab} \R^{cd} \we \S_{cd} \approx 0 \lb{eqmo1} \E

\B \frac{2k}{\sqrt{-g}}\  \frac{\d \L}{\d \th^a}  = \R^{bc} \we \S_{bca} \approx 0 \lb{eqmo2} \E

\B \frac{2k}{\sqrt{-g}}\  \frac{\d \L}{\d \w^a_{\ b}} = \Mb^{bc} \we \S_{ac} + \X^c \we \S_{aec} g^{eb} := \frac{1}{\sqrt{-g}} \cex \left(\sqrt{-g} g^{bc} \S_{ac} \right) \approx 0 \lb{eqmo3} \E

where we used the definitions: $\Mb^{ab} := \M^{ab}-g^{ab} \frac{\M}{2}$ and $\M := \M^{ab} g_{ab}$.
We will analyse the meaning of these equations of motion and their relation with usual formulations of General Relativity in the next subsections.

\subsection{Local symmetries and associated superpotentials}

The gravitational Lagrangian \ref{actgrav} has three distinct gauge symmetries. Let us apply the cascade machinery for each one. 

\VS

\VS

  1) $\mathit{The\ local\ ``frame\ choice"\ freedom}$ which is of the Yang-Mills type.

Under a local infinitesimal linear transformation $\Lambda^a_{\ b}(x) = \d^a_{\ b} + \l^a_{\ b}(x)$,  the variations of $\th^a$ and $\w^a_{\ b}$ are given by:

 $\d_\l \th^a (x) = \l^a_{\ b} (x) \th^b (x) $

 $\d_\l \w^a_{\ b} (x) = \l^a_{\ c} (x) \w^c_{\ b} (x) - \l^c_{\ b} (x) \w^a_{\ c} (x)  - \ex \l^a_{\ b} (x)$

It is easy to check that the Lagrangian \ref{actgrav} remains invariant under this transformation, i.e.  $\mathbf{\d_\l L}$  $=0$. 
Since $\L$ depends on derivatives of the $\w^a_{\ b}$ field only (there is no $\ex \th^a$ or  $\ex g^{ab}$ terms in \ref{actgrav}), equation \ref{cur1} of the general discussion is in this case simply

\B   \d_\l \L - \ex \left(\d_\l \w^a_{\ b} \we \frac{\p \L}{\p \ex  \w^a{_b}}  \right)  \approx 0 \lb{invar} \E

Which becomes after substitution

\B \ex \J_\l \approx 0 \lb{avoir}\E

Where 

$\J_\l := \l^a_{\ b} \J^b_{\ a} + \ex \l^a_{\ b} \we \U^b_{\ a}$

$\J^b_{\ a} :=  \frac{\sqrt{-g}}{2k} \left( - \w^b_{\ c}  \we \S_{ad} g^{dc}  + \w^c_{\ a} \we \S_{cd} g^{db} \right)$

$\U^b_{\ a} :=  \frac{\sqrt{-g}}{2k} \S_{ad} g^{db}$


The cascade equations are derived as before: we just impose the fact that $\l^a_{\ b}$ and its derivatives are arbitrary:

\B  \ex\l^a_{\ b} \we \left[\J^b_{\ a} \approx \ex \U^b_{\ a} \right] \lb{jcas2}\E

\B \l^a_{\ b}  \left[ \ex \J^b_{\ a} \approx 0 \right] \E

Note that the 'zero'$^{th}$ cascade equation which in component language was the antisymmetry of the superpotential here is automatically satisfied because we use differential forms.

These equations are nothing other than the equations of motion of $ \w^a_{\ b}$ (see equation \ref{eqmo3}) as in the Yang-Mills case. 

Now, if we use the abelian cascade trick, we can rewrite the total current $\J_\l$ as a superpotential. The result is obvious:

\B  \J_\l \approx \ex \U_\l \lb{glcas} \E

Where $\U_\l := \l^a_{\ b} \U^b_{\ a}$.

Later we will analyse the connection of this theory with ordinary General Relativity and we will see that this superpotential appears in the conservation of angular momentum. But before that, let us pursue the discussion of local symmetries.

\VS

  2) $\mathit{The\ diffeomorphism\ invariance}$. Under an infinitesimal local reparameterisation $x^\r \rightarrow x^\r + \xi^\r (x)$, the variation of the Lagrangian is given by its Lie derivative $\li_{\xi} \L$ along $\xi^\r$. Now $\ex\L= 0$ because $\L$  is a top form and $\li_{\xi} =\ex \cdot \i_{\xi}  +  \i_{\xi} \cdot \ex$ so we see that the variation of the Lagrangian is a total derivative, i.e. $\li_{\xi} \L=\ex \cdot \i_{\xi} \L$. 
In addition, $\d \w^a_{\ b} = \li_{\xi} \w^a_{\ b}$ and so the conservation equation becomes

\B  \ex \left( \i_{\xib}\L - \frac{\sqrt{-g}}{2k} \left[ (\ex \cdot \i_{\xi}  +  \i_{\xi} \cdot \ex )\w^a_{\ b} \right]  \we g^{bd} \S_{ad} \right) \approx 0 \lb{der1}\E

\B  \Leftrightarrow \ex \left( \xi^\r \i_\r \L - \frac{\sqrt{-g}}{2k} \left[ (\ex \xi^\r \cdot \i_\r  +\xi^\r  \li_\r  )\w^a_{\ b}\right]  \we g^{bd} \S_{ad}\right) \approx 0 \lb{der2}\E

Where for notational convenience we defined $\i_\r := \i_{\frac{\p}{\p_\r}}  $ and $\li_\r := \li_{\frac{\p}{\p_\r}}$.

We can then separate out the factors of $\xi^\r$ and $\ex\xi^\r$ and, define the (D-1) and (D-2) forms respectively, 

\VS

$\t_{\r} := \i_\r \L - \frac{\sqrt{-g}}{2k} \left( \li_\r  \w^a_{\ b} \right) \we  g^{bd} \S_{ad} $

$\s_{\r} := - \frac{\sqrt{-g}}{2k}  \left( \i_\r \w^{ad}\right) \S_{ad}$

\VS

to finally get

\B \ex \J_\xi :=  \ex ( \xi^{\r} \t_{\r} + \ex\xi^{\r} \we \s_{\r}) \approx 0 \lb{totder} \E

Again, the factors in front of $\xi^{\r}$ and $\ex \xi^{\r}$ must vanish separately :

\B \ex \xi^{\r} \we \left[ \t_{\r} - \ex \s_{\r}\right] \approx 0 \lb{1cas2}\E

\B \xi^{\r} \left[ \ex \t_\r \right] \approx 0 \lb{1cas3}\E

The abelian cascade gives us:

\B \J_\xi \approx \ex \U_\xi \lb{difcas}\E

where $\U_\xi := \xi^\r \s_\r$.

The cascade equations of $gl(D,\Re)$ gauge invariance was just a rewriting of the equations of motion for $\w^a_{\ b}$. In the next subsection we will explain how to recover General Relativity, the cascade equations \ref{1cas2} and \ref{1cas3} will encode some  Einstein equations derived by variation of $\th^a$ and $g^{ab}$ (see equations \ref{eqmo1} and \ref{eqmo2}). We will see that in section 4.3.

 Before that, let us analyse the third symmetry of the Lagrangian \ref{actgrav}.

\VS

\VS

  3) $\mathit{The\ Projective\ Symmetry}$ is a generalisation of what is called the $\l$-Einstein symmetry \c{HeM}. It is very easy to check that \ref{actgrav} is invariant under

$\d_\k \w^a_{\ b} = \kb \d^a_{\ b}$

$\d_\k \th^a =\d_\k g^{ab} = 0$

where $\kb$ is an arbitrary 1-form. For completeness we note that under such a variation, the curvature varies as $\d_\k \R^a_{\ b} = \ex \kb \d^a_{\ b}$ and then if $\kb = \ex \l$ (the $\l$-Einstein symmetry) it stays invariant. 
More generally the projective symmetry follows from the elimination of the diagonal $\Re$ subgroup of $GL(D,\Re)$ by our choice of dynamics.

Again, the conservation equation for this symmetry reads

\B \ex \J_{\kb} := \ex \left(\kb \we \frac{\p \L}{\p \ex \w^a_{\ a}}  \right) \approx 0 \E

If we follow the cascade we obtain as first equation that

$$\ex \kb \we \frac{\p \L}{\p \ex \w^a_{\ a}}  \approx 0$$

And the Noether current has to vanish. This is due to the vanishing of the  local superpotential, or in other words to the fact that we have a local symmetry without any propagating field that transforms as a derivative of the parameter (in particular there is no gauge potential for this local symmetry). This is a theorem:

\VS

\underline{Theorem} 1: If propagating fields have transformation rules that do not contain derivatives of the local $\mathit{arbitrary}$ parameter and if the variation of the Lagrangian can be written as the divergence of a surface term with that same property, then the Noether current has to vanish.

\VS

There are many well known examples of this fact, for instance $kappa$ symmetry or Weyl symmetry of string theory. 
The fact that the parameter of the symmetry is unconstrained is fundamental for the theorem. As we will see in the last section for the case of (non)homentropic fluids, a constraint on the parameter allows conserved charges, even an infinite number of them in even space dimension. Note that the theorem may break down for diffeomorphisms symmetry if fields have spin because of the derivative terms or because of surface terms as in Einstein Lagrangian.

Although this projective symmetry gives no contribution to the total Noether current or to the superpotentials, its existence is crucial if we want to identify somehow the affine gauge theory with General Relativity. In fact the Einstein theory will be recovered as a special gauge choice as we will see in the following subsection.

\bigskip

\subsection{Noether identities and integrability condition for gravitation}

\VS

\underline{The Noether Identities}

First of all we would like to add some precisions about the equations of motion \ref{eqmo1}, \ref{eqmo2} and \ref{eqmo3}. The Noether identities due to the gauge symmetry are relations between the equations of motion. For instance if we use equation \ref{2no} for the $gl(D,\Re)$ symmetry we obtain after some rearrangement that 

\B \cex \left( \frac{\d \L}{\d \w^a_{\ b}} \right)+  \th^b \we \frac{\d \L}{\d \th^a}  + 2 g^{eb} \frac{\d \L}{\d g^{ae}}   = 0 \lb{noid} \E

Using now \ref{eqmo3} and the Bianchi identities \ref{bian1}, \ref{bian2} and \ref{bian3} we easily find that \ref{noid} implies (after lowering the $^b$ index)

\B \R^c_{\ a} \we \S_{bc} - \R^c_{\ b} \we \S_{ac} +  \th_b \we \frac{\d \L}{\d \th^a}  + 2\frac{\d \L}{\d g^{ab}}  = \left( \R_{bc} + \R_{cb} \right)\we \S^c_{\ a} \lb{idnocc} \E

The Einstein theory has zero nonmetricity. In that case, the Bianchi identity \ref{bian3} implies that the symmetric part of the curvature has to vanish $R^{(ab)}=R_{(ab)}=0$, and so the r.h.s of \ref{idnocc} can be set to zero.
Using this vanishing condition the first pair of terms of this equation is completely antisymmetric in $ab$ but the last term is completely symmetric. 
The conclusion of the Noether identity is then that the equations of motion of the metric are identical to the symmetric part of those of the moving frame when the nonmetricity vanishes. Of course these equations are nothing more than the vacuum Einstein equations after we identify $\R^a_{\ b}$ with the Riemann tensor $\R^a_{\ b} (g)$. 
It is also interesting  to discuss the antisymmetric part of \ref{idnocc}. Again assuming metricity (and so the r.h.s. of \ref{idnocc} vanishes), the expression $\th_{b[} \we \frac{\d \L}{\d \th^{a]}}$ is actually a combination of the torsion Bianchi identity \ref{bian2}. It is an easy exercise to use the invariance of $\e_{a_1 \cdots a_D}$ under $sl(D,\Re)$ (using again metricity) to prove this fact.

The Noether identity due to the projective symmetry is rather simple:

\B \frac{\d \L}{\d \w^a_{\ a}} = 0 \lb{nopro} \E

Its meaning will be explained in the following. Before that, we would like to analyse the Noether identities due to the diffeomorphism invariance. 

\VS

``\underline{The Einstein equations are conservation laws}"

The purpose is to reanalyse with our technology the following theorem:

\VS

\underline{Theorem} 2: The following three affirmations are equivalent:

1- Spacetime is Ricci-flat with null torsion and null nonmetricity in the Einstein gauge.

2- $\ex \t_\r = 0\ \forall _\r$.

3- $\t_\r = \ex \s_\r\ \forall _\r$.

\VS

A similar theorem was obtained some time ago by Dubois-Violette and Madore \c{Du} but considering neither variational principle nor gauge symmetry.

\VS

Proof:

\hspace{.3in} $\bullet$ $2 \Rightarrow 1$
 
First note that we can rewrite the on-shell equation \ref{1cas3} as an off-shell equation by use of the formula \ref{idcas1} of the general discussion (section 2.1) applied to the diffeomorphism invariance. The result is:

\B \ex \t_\r = \li_\r \left( g^{ab} \right)\  \frac{\d \L}{\d g^{ab}} + \li_\r \left( \th^a \right) \we \frac{\d \L}{\d \th^a} +  \li_\r \left( \w^a_{\ b} \right) \we \frac{\d \L}{\d \w^a_{\ b}} \lb{dv}\E

In addition to that, the Noether identity (equation \ref{2no}) due to the diffeomorphism invariance gives:

\B \li_\r \left( g^{ab} \right)\  \frac{\d \L}{\d g^{ab}} + \li_\r \left( \th^a \right) \we \frac{\d \L}{\d \th^a} +  \li_\r \left( \w^a_{\ b} \right) \frac{\d \L}{\d \w^a_{\ b}} = \ex \left( \i_\r \left( \th^a \right)\  \frac{\d \L}{\d \th^a} +  \i_\r \left( \w^a_{\ b} \right)\ \frac{\d \L}{\d \w^a_{\ b}} \right) \E

Then equation \ref{dv} just becomes

\B \ex \t_\r = \ex \left( \i_\r \left( \th^a \right)\  \frac{\d \L}{\d \th^a} +  \i_\r \left( \w^a_{\ b} \right)\ \frac{\d \L}{\d \w^a_{\ b}} \right) \lb{dv2} \E

Note that the equations of motion of the theory (see explicit computations, equations \ref{eqmo1}, \ref{eqmo2} and \ref{eqmo3}) transform as $gl(D, \Re)$-tensors. So, the only part of the r.h.s. of \ref{dv2} which is not a $gl(D, \Re)$ scalar is the last term which is proportional to 
$\i_\r \left( \w^a_{\ b} \right)$. 
Thus, if we suppose that the l.h.s. is zero, this last term has to vanish identically (otherwise a $gl(D, \Re)$ gauge transformation parametrized by $\l^a_{\ b}$ would generate the non vanishing term $\i_\r \left( \ex \l^a_{\ b} \right)\ \frac{\d \L}{\d \w^a_{\ b}}$), so we must have that $\frac{\d \L}{\d \w^a_{\ b}} = 0$.
 We will admit now that this last equation implies that the torsion $\X^a$ and the nonmetricity $\M^{ab}$ have to vanish in what will be called the Einstein gauge. This important statement will be proved in the following discussion: ``fixing the projective symmetry"  of section 4.4.
If we use all this in equation \ref{dv2} we obtain that affirmation 2 implies that

\B \ex \left( \theta^a_\r \  \frac{\d \L}{\d \th^a} \right) = \cex \left( \theta^a_\r \  \frac{\d \L}{\d \th^a} \right) = 0 \E

Now we can use equation \ref{eqmo2} for $\frac{\d \L}{\d \th^a}$, the Bianchi identity \ref{bian1} for $\R^a_{\ b}$ and the vanishing of torsion and nonmetricity to rewrite equation \ref{dv2} as:

\B - \R^{bc} \we \S_{bca} \we \cex \theta^a_\r = 0 \lb{eqsimple} \E

This holds for arbitrary coordinate frames and so the Einstein equations are satisfied. 

\VS

\hspace{.3in} $\bullet$ $3 \Rightarrow 2$

This statement is obvious. 

\VS

\hspace{.3in} $\bullet$ $1 \Rightarrow 3$

We can use equation \ref{idcas2} of the general discussion (section 2.1) applied to the diffeomorphism invariance to obtain:

\B \t_\r - \ex \s_\r = \theta^a_\r \  \frac{\d \L}{\d \th^a} +  \omega^a_{\r b} \ \frac{\d \L}{\d \w^a_{\ b}} \E

If the torsion and the nonmetricity vanish and the Einstein equations are satisfied the r.h.s. is obviouly zero, which implies affirmation $3$.

\VS

What we have obtained is thus just a new way to rewrite the Einstein equations in D dimensions for vanishing torsion and nonmetricity. The theorem of Dubois-Violette and Madore \c{Du} uses objects that are not our $\t_\r$ and $\s_\r$ and were just postulated in $D$ dimensions after the work of Sparling \c{Sp} and Nester-Witten  \c{Wi} \c{Ne} in $4$ dimensions.
To be more precise, these objects were first defined directly on the bundle of $\mathit{orthonormal}$ frames (i.e. with gauge group $so(D, \Re)$) and later on the linear frame bundle $L(M)$ by Frauendiener \c{Fr}. 
Our formulas are based on a variational principle and on symmetry arguments and 
so cannot be obtained on the bundle. We have to choose an arbitrary section 
first and then the computations can begin. We can then hope to pullback the 
quantities of Dubois-Violette and Madore along this arbitrary section. This 
will be done in section 5.1.

\VS

\subsection{Gauge fixing and equivalence with Palatini and Cartan-Weyl actions}

``\underline {Gauge fixing the Projective Symmetry}"

\VS

At the end, we would like to recover the basic statements of Einstein theory which are the vanishing of torsion and nonmetricity. The first one imposes on our theory $\frac{D^2 (D-1)}{2}$ constraints while the second imposes $\frac{D^2(D+1)}{2}$ more constraints, so we just need $D^3$ constraints. One could have hoped that these constraints would be given by the naively $D^3$ equations of motion of $\w^a_{\ b}$ \ref{eqmo3}. However the projective symmetry with its Noether identity is telling us that the $\w^a_{\ b}$ give only $D^3-D$ independent equations ($D$ of them are $\mathit{identically}$ zero, equation \ref{nopro}). After some algebraic work we derive from equations \ref{eqmo3} that their general solutions are given by:

\VS

$\X^a \approx \La \we \th^a$

$\Mb^{ab} \approx (2-D) \La g^{ab}$

\VS

Where $\La$ is an arbitrary (undetermined by the equations of motion) one form. In fact it is just proportional to the trace of the nonmetricity, namely $\La= \frac{\M}{2D}$. Thus we obtain only ($D^3-D$) independent equations which are not enough by themselves to get a null torsion and nonmetricity. In other words, the trace of the nonmetricity will be a free field, not fixed by the equation of motion of $\w^a_{\ b}$. However, if we fix the projective symmetry in the Einstein gauge defined by:

\VS 

$\tilde{\w}^a_{\ b} = \w^a_{\ b} - \La \d^a_{\ b}$

\VS

then we obtain the wanted result

$\Rightarrow \tilde{\X}^a = \tilde{\Mb}^{ab} = 0 $ and so $\tilde{\M}^{ab} = 0$.

We will henceforth that the Einstein gauge is used. 

\VS

``\underline{Gauge fixing the $gl(D,\Re)$ Symmetry}"

\VS

We will use the following three gauge choices:

\VS

- $\mathit{the\ holonomic\ or\ coordinate\ gauge}$: $\theta^{\ a}_\m =\d^{\ a}_\m$ (or in differential form notation, $s^* (\th^a) = \ex x^a$). Here we will be able to use indifferently greek (for the curved base manifold) or latin indices ($gl(D,\Re)$ ones).

A frame choice makes use of all the $D^2$ degrees of freedom of $gl(D,\Re)$ but the theory still has the diffeomorphism symmetry. The above gauge fixing has to be preserved under a Lie-derivative transformation, and if it is not the case, has to be compensated by a $gl(D,\Re)$ transformation. In fact we see that

\VS

$ \li_{\xi} \theta^{\ a}_\m|_{\theta =\d} = \xi^\r \p_\r \d^{\ a}_\m + \p_\m \xi^\r \d^{\ a}_\r = \p_\m \xi^a $

$ \l ^a_{\ b} \theta^{\ b}_\m|_{\theta = \d } = \l ^a_{\ \m}$

\VS

Thus, the diffeomorphism transformation which preserves the coordinate gauge is a Lie derivation of parameter $\xi^\r$ in addition to a $gl(D,\Re)$ rotation of parameter $\l ^a_{\ \m}=-\p_\m \xi^a$. Note that with this we recover for example the usual Lie derivative formula for the metric: $\d g^{ab} = \xi^\r \p_\r g^{ab} + \l^a_{\ e} g^{eb} + \l^b_{\ e} g^{ae} = \xi^\r \p_\r g^{ab} -\p_e \xi^a g^{eb} -\p_e \xi^b g^{ae}$. The same is true for the connection field which now has to be identified with the Christoffel symbols (the Einstein gauge is used), $\w^a_{\ b} = \G^a_{\m b} \ex x^\m$ (the non tensorial part of the connection transformation is given by the $-\ex \l^a_{\ b} = \p_\m \p_{\ b} \xi^a \ex x^\m$ term).

This will be used in the next section to study the gravitational superpotentials.

We would like now to reduce our theory to the Palatini formalism: in the Einstein gauge, the equations of motion of $\G^a_{[\m b]}$ will imply the vanishing the torsion. The point is that in a coordinate frame, the vanishing of torsion $\mathit{is}$ the vanishing of $\G^\r_{[\m \n]}$ (remember that $\X^a=\ex \db^a + \w^a_{\ b} \we \db^b = \G^a_{[\m b]} \ex x^\m \we \ex x^b = 0$). 

If we eliminate the $\G^\r_{[\m \n]}$ fileds from the Lagrangian \ref{actgrav} 
 we recover the Palatini first order formulation of gravity depending on the fields $\G^\r_{(\m \n)}$ and $g^{\m\n}$.
The equations of motion of the first give the metricity condition (i.e. the equation which gives $\G$ in terms of $g$ and its derivatives) and those of the second the Einstein equations.
More precisely the equations of motion of the $\th^a$ field which is eliminated by going to a coordinate frame must be recovered as equations of motion of $\G^\r_{(\m \n)}$ and $g^{\m\n}$. 
We have seen in the study of Noether identities of $gl(D,\Re)$ how the symmetric and antisymmetric parts of \ref{idnocc} are indeed deducible from them when torsion vanishes.

\VS

- $\mathit{the\ orthogonal\ gauge}$: $g^{ab} = \eta^{ab}$ ($\eta^{ab}$ is the usual flat metric).

This gauge condition fixes only $\frac{D(D+1)}{2}$ of the $D^2$ degres of freedom allowed by $gl(D,\Re)$. The remaining symmetries are now the Lie derivative diffeomorphism invariance and the local $so(D,\Re)$ Lorentz invariance, parametrized now by an infinitesimal antisymmetric tensor $\e^a_{\ b}$, $\e^{ab}=\e^a_{\ e} \eta^{eb} = \e^{[ab]}$. The easy part here is that we do not need to modify these symmetries by a compensating ``symmetric $gl(D,\Re)$ rotation" because the gauge choice is automatically preserved:

\B \li_{\xi} g^{ab}|_{g= \eta} = \xi^\r \p_\r \eta^{ab}=0 \E

\B \left( \e ^a_{\ e} g^{eb} + \e ^b_{\ e} g^{ea} \right)|_{ g = \eta}=\e^{ab} + \e^{ba} =0 \E

The relation with the usual tetrad or vielbein (Cartan Weyl) theory is as follows: In the Einstein gauge 
the equations of motion of $\w^{(ab)}$ imply that the nonmetricity has to vanish. Now the vanishing of the nonmetricity $\mathit{is}$ the vanishing of $\w^{(ab)}$ when the orthogonal gauge is used (remember that $\M^{ab}=\ex \eta^{ab} +\w^{ab} + \w^{ba} = 2 \w^{(ab)} = 0$).

There is a nice way to understand this: let us decompose $\w^{ab}$ in its irreducible parts $\w^{ab}_s=\w^{(ab)}-\w^c_{\ c} \frac{g^{ab}}{D}$ (symmetric traceless), $\w^{ab}_{A}=\w^{[ab]}$ (antisymmetric) and $\w^{ab}_{T} = \w^c_{\ c} \frac{g^{ab}}{D}$ (trace). The curvature can also be decomposed in the same way $\R^{ab}_S= \ex \w^{ab}_S + \w_{S\ c}^{\ a} \we \w_{A}^{cb} + \w_{A\ c}^{\ a} \we \w_S^{cb}$, $\R^{ab}_{A}= \ex \w^{ab}_{A} + \w_{A\ c}^{\ a} \we \w_{A}^{cb} + \w_{S\ c}^{\ a} \we \w_S^{cb}$ and $\R^{ab}_{T} = \ex \w_{T}^{ab}$. 
Note that only the skew part of the curvature ($\R^{ab}_{A}$) contributes to the Lagrangian \ref{actgrav} (due to the contraction of the curvature with the antisymmetric tensor $\S_{ab}$) 
and so $\w_{T}^{ab}$ completely decouples, the Einstein gauge corresponds to 
set it to zero. The symmetric $\w^{ab}_S$ has no kinetic term and its equation 
of motion equals it to zero. We recover in that way the above conclusion.
 
So the field $\w^{ab}_S$ can be eliminated from the Lagrangian \ref{actgrav} to obtain the first order tetrad formulation of gravity, i.e. a theory which depends on the fields $\w^{[ab]}$ and $\th^a$ where the equations of motion of the first give the null torsion equation and of the second the Einstein equations.
Note that the equations of motion of $g^{ab}$ have not been lost by the gauge choice $g^{ab}=\eta^{ab}$ as we discuss at the beginning of section 4.3. They are indeed in those of the form $\th^a$ (see equation \ref{idnocc}).

\VS

- $\mathit{the\ arbitrary\ fixed\ frame}$: $\theta^{\ a}_\m = \bar{\theta}^{\ a}_\m$.

This case is slightly more complicated than the previous ones. Now the frame is chosen to be an arbitrary $x$-dependent frame $\bar{\theta}^{\ a}_\m (x)$. Again, all the $gl(D,\Re)$ gauge invariance has been used. The modified remaining diffeomorphism invariance can be computed as in the previous example:

\VS

$ \li_{\xi} \theta^{\ a}_\m|_{\theta =\bar{\theta}} = \xi^\r \p_\r \bar{\theta}^{\ a}_\m + \p_\m \xi^\r \bar{\theta}^{\ a}_\r $

$ \l ^a_{\ b} \theta^{\ b}_\m|_{\theta =\bar{\theta}} = \l ^a_{\ b}\bar{\theta}^{\ b}_\m$

\VS
Then the remaining symmetry combines a Lie derivation of parameter $\xi^\r$ plus a $gl(D,\Re)$ rotation of parameter $\l ^a_{\ b} = 
-\bar{\theta}_{\ b}^\m ( \xi^\r \p_\r \bar{\theta}^{\ a}_\m + \p_\m \xi^\r \bar{\theta}^{\ a}_\r)$.

\bigskip

\subsection{Comparison of superpotentials}

The purpose of this subsection is to collect all the above results to recover some of the well known superpotentials. The connection with the second order formulation of section 3 will also be established.

We learned in previous sections that the local $gl(D,\Re)$ invariance implies the existence of a total Noether current $\J_\l$ together with an associated superpotential $\U_\l$ (equation \ref{glcas}) both $\l^a_{\ b} (x)$ dependent. The same is true for the local diffeomorphism invariance (parametrized by $\xib (x)$) , with corresponding quantities $\J_\xi$ and $\U_\xi$ (equation \ref{difcas}). The $\kb$ local symmetry had vanishing currents for the reason already explained at the end of section 4.2. With all this we can construct a big Noether current:

\B \J_{\xi,\l} \approx \ex \U_{\xi,\l} \lb{bigno} \E

where

\VS

$\J_{\xi,\l} := \J_\xi + \J_\l = \xi^\r \t_\r + \ex \xi^\r \we \s_\r + \l^a_{\ b} \J^b_{\ a} + \ex \l^a_{\ b} \we \U^b_{\ a}$

$\U_{\xi,\l} := \U_\xi + \U_\l =\xi^\r \s_\r + \l^a_{\ b} \U^b_{\ a}$

\VS

Now, if we use an holonomic section $s$ of $L(M)$ (in other words if we use a coordinate frame) , i.e. $\th^a = \ex x^a$, then:

\VS

\hspace{0.5in} $\bullet$ the dual of the ($D-2$)-form $\s_\r$ is what we called 
the M\o ller superpotential in  equation \ref{mol2}. In fact, M\o ller 
multiplied it by a factor of two in a desesperate attempt to gain weight \c{Mo}.  From the definition of $\s_\r$ (equation \ref{totder}) we have that:

$\s_\r = -\frac{\sqrt{-g}}{2k} \omega_\r^{ab} s^*(\S_{ab})$

$\hspace{.47in}          = -\frac{\sqrt{-g}}{2k} \omega_{\r c}^{[a}\ g^{b]c}\  s^*(\S_{ab})$

\VS

and, using the fact that $s$ is a coordinate section, $\w^a_{bc} =\G^a_{bc}$ with the definition \ref{mol2} we obtain,

\VS

$s^*(\s_\r) = \frac{1}{2}\ _MU_\r^{[ab]} s^*(\S_{ab})$

$\hspace{.47in}  =\frac{1}{2} \ _MU_\r^{[ab]} \frac{1}{(D-2)!} \e_{ab c_3...c_D} \ex x^{c_3} \we ...\we \ex x^{c_D}$

In components, the dual 2-form of this is just $\ _MU_\r^{[\m\n]}$ (the $\frac{1}{2}$ factor is killed as usual by the $2$ coming from the $\e_{ab c_3...c_D} \e^{\m\n c_3...c_D}$ contraction). 

\VS

We also find that

\VS

$s^*(\ex\s_\r) = \ex s^*(\s_\r) = \frac{1}{2} \p_\a  \ _MU_\r^{[\m\n]} \ex x^\a \we \frac{1}{(D-2)!} \e_{\m \n c_3...c_D} \ex x^{c_3} \we ...\we \ex x^{c_D}$

$ \hspace{.6in}            = \p_\n\ _MU_\r^{[\m\n ]} \frac{1}{(D-1)!} \e_{\m c_2...c_D} \ex x^{c_2} \we ...\we \ex x^{c_D} $

\VS
\hspace{0.5in} $\bullet$ the dual of the ($D-1$)-form $\t_\r$ is just the canonical energy momentum complex (of equations \ref{2cas3} and \ref{2cas4}) or one half of M\o ller's original gravitational energy-momentum pseudotensor. Let us recall that $\L = \frac{\sqrt{-g}}{2k} R \S$ where $R$ is the scalar curvature and $\S=\frac{1}{D!} \e_{c_1...c_D} \ex x^{c_1} \we ...\we \ex x^{c_D} $ the volume form, we obtain from the definition of $\t_\r$:

\VS

$\t_\r =\frac{\sqrt{-g}}{2k} R s^*(i_\r\S) -\frac{\sqrt{-g}}{2k} \p_\r \omega_{\m b}^{[a}  g^{d]b} s^*(\th^\m \we \S_{ad})$

$ \hspace{.5in} =   \frac{\sqrt{-g}}{2k} R s^*(\S_\r) - 2 \frac{\sqrt{-g}}{2k} \p_\r \omega_{\a\b}^{[\a}  g^{\g ]\b} s^*(\S_\g)$

$ \hspace{.5in} =\ _MT_\r^{\ \m} \frac{1}{(D-1)!} \e_{\m c_2...c_D} \ex x^{c_2} \we ...\we \ex x^{c_D}$

\VS

\hspace{0.5in} $\bullet$ the dual of the ($D-2$)-form $\U_{\xi,\l}$ with $\l^a_{\ b}=-\p_b \xi^a$ (see section 4.4)  is just the Komar superpotential \ref{kom}. To show that we first use (from equation \ref{avoir}) that:

\VS

$\U^b_{\ a} = \frac{\sqrt{-g}}{2k} g^{bd} s^*(\S_{ad})$

$ \hspace{.5in} = \frac{1}{2} W_a^{\ b[dc]} s^*(\S_{cd}) $

Where $W_a^{\ b[dc]}$ has been defined in \ref{wdef}. We thus see that

$\U_{\xi,-\p \xi} = \xi^\r \s_\r - \p_a \xi^b s^*(U^b_{\ a})$

$ \hspace{.75in} = \frac{1}{2} \left(\ _MU_\r^{[\m\n ]} \xi^\r + W_\r^{\ \a[\m\n]} \p_\a \xi^\r \right) s^*(\S_{\m\n}) $

$ \hspace{.75in} =  \frac{1}{2} \ _KU_\xi^{[\m\n ]} s^*(\S_{\m\n})$

\VS

\hspace{0.5in} $\bullet$ the dual of the ($D-1$)-form $\J_{\xi,\l}$ with $\l^a_{\ b}=-\p_b \xi^a$ is just the total Noether current, equation \ref{epcas} or \ref{eqbase}. To simplify the analysis and obtain a better understanding of what is going on, let us first make the following comment: in the second order formalism the definition of $\G^\m_{\n\r}$ in terms of the metric and its derivatives
is assumed, in our language, this means that the equations of motion of 
$\w^a_{\ b}$ have been used (in the Einstein gauge). This means that equation \ref{jcas2} becomes an identity and then the big Noether current (definition \ref{bigno}) specializes for $\l=-\p \xi$:

\VS

$\J_{\xi,-\p \xi} := \J_\xi + \J_{\l=-\p \xi} = \xi^\r \t_\r + \ex \xi^\r \we \s_\r - \p_b \xi^a  \ex \U^b_{\ a} - \ex \p_b \xi^a \we \U^b_{\ a}$

\VS

We then obtain:

$\J_{\xi,-\p \xi} = \xi^\r \ _MT^{\ \m}_\r s^*(\S_\m) + \left( \p_\a \xi^\r \ _MU_\r^{[\m\n]} + \p_b \xi^a  \p_\a  W_a^{\ b[\m\n]} \right.$

$\hspace{1in} \left. + \p_\a \p_b \xi^a  W_a^{\ b[\m\n]} \right) \frac{1}{2} \ex x^\a \we s^*(\S_{\m\n}) $

\B = \left[ \xi^\r \ _MT^{\ \m}_\r + \p_\n \xi^\r \left(\ _MU_\r^{[\m\n]} +\p_\d W_\r^{\ \n\m\d} \right) + \p_\d \p_\n \xi^\r  W_\r^{\ \n\m\d} \right] s^*(\S_\m) \lb{spe} \E

\VS

Now we are in a position to understand the objects found in the second order formulation: the second term in \ref{spe} is just what was called the canonical spin complex $U_\r^{\ \m\n}$ and the last term in \ref{spe} symmetrized on the $(\n \d)$ indices of course is just $V_\r^{\ \m\n\d}$ of section 3.1. The somewhat mysterious relation \ref{luck}, and the consequent non antisymmetry of the spin complex becomes natural in this framework. The completely antisymmetric contribution comes from the diffeomorphism but another term comes from a $gl(D,\Re)$ compensating transformation. We can understand in a similar way equation \ref{luck2}. We point out now that the difference between the canonical spin complex and the M\o ller superpotential is not a physical observable since the conserved charge is always obtained by its integral over a $\mathit{closed}$ boundary (for instance, see equation \ref{loccha}).

\VS

We will not analyse in more details the results of the previous discussion for orthonormal sections. We just want to point that for an arbitrary fixed section, we obtain again the Komar superpotential if we use the modified $\l^a_{\ b}$ given at the end of section 4.4 also for that case.

\VS

Finally, we have obtained

- An explicit derivation of superpotentials from first principles (gauge invariance) and a general formula \ref{cur2}.

- The Affine formulation gives the superpotentials in a clearer geometrical way, if reduces to the usual formulations of gravity by using some equations of motion and fixing the Einstein gauge. This is why we will study it further in section 5.

- Remember that the abelian cascade defined in section 2.1 is used to obtain a parameter dependent superpotential. When we used it before compensating for  the choice of a coordinate 
section we obtained an explicitly antisymmetric superpotential, $\U_\xi$, 
(which  gives the Komar superpotential when the compensation is taken into 
account). When we used it directly in the $2^{nd}$ order formulation we got a 
non antisymmetric one (see equation \ref{esup}), which has been shown to be 
identical to the first one up to a divergence term \ref{luck2}.
 It seems that more geometrical objects are obtained when the abelian cascade is used in the $gl(D,\Re)$ formalism. This is the method we will use in the following.

- About the mysterious one half factor between the superpotentials we found 
from our general formula \ref{cur2} and the ad-hoc definitions given in the 
early days by M\o ller \c{Mo} and Komar \c{Ko}, the point is that, as we will 
show in the next section, the variational principle and the Lagrangian \ref{actgrav} used in this section are not compatible with standard asymptoticaly flat solutions of general relativity as for example the Kerr solution. It then does not make sense to use the corresponding superpotentials
 to compute for example the conserved mass or angular momentum. If we in fact compute the conserved charge associated with the constant asymptotic Killing time like vector for the Schwarzschild solution, we obtain with our definitions $\frac{m}{2}$. We understand now why M\o ller and then Komar doubled the above expressions in the hope to obtain correct superpotentials. As we will see in next section, to correct the above anomaly we will need to add a surface term to our first Lagrangian \ref{actgrav}. The associated superpotentials will be now those of Freud \c{vF} and Katz \c{Ka} (actually, a non background version of it). They are comparable with the ADM formula (see for example \c{RT}) of the Hamiltonian formalism.

\bigskip

\section{ Surface terms, well posed variational principles and physical charges}

\VS

\subsection{Physical Lagrangian, modified superpotentials}

As we discussed in the Yang-Mills situation (section 2.2) the choice of boundary conditions will specify the surface term we have to add to the Lagrangian of the theory. For instance for the Lagrangian \ref{actgrav}, the variational principle implies that (equation \ref{varcon} in the Yang-Mills case)

\B \int_{\p M} \d  \w^a_{\ b} \we \frac{\p \L}{\p\ex \w^a_{\ b}}   \lb{varcon2}\E

has to vanish.

Later, we would like to impose the vanishing of metricity and torsion which, as we know, allows us to rewrite the $\w^a_{\ b}$ (or the $\G^a_{\m b}$ in more familiar notation) in terms of the metric and its derivatives. But the condition $\d  \w^a_{\ b} = 0$ on boundaries is a Neumann type boundary condition (for the metric). If we look for conserved charges, then we would like instead to impose Dirichlet boundary conditions. For that, we have to eliminate all the derivatives of $\w^a_{\ b}$ from the Lagrangian \ref{actgrav} by adding a surface term. The most obvious way to do that (and the only one, see \c{CN}) is to define:

\VS

$ \hspace{0.5in} \hat{\L} = \L - \ex\Sg $ 

\B = \frac{\sqrt{-g}}{2k} \w^a_{\ c} \we \w^{cb} \we \S_{ab} + \frac{1}{2k}  \w^a_{\ e} \we \ex( \sqrt{-g} g^{eb} \S_{ab}) \lb{Ein}\E

$\hspace{0.7in} = -\frac{\sqrt{-g}}{2k} \w^a_{\ c} \we \w^{cb} \we \S_{ab} + \frac{1}{2k}  \w^a_{\ e} \we \cex( \sqrt{-g} g^{eb} \S_{ab}) $

\VS

Where $\Sg := \frac{1}{2k} \w^a_{\ e} \we \sqrt{-g}  g^{eb}  \S_{ab}$. Note that this Lagrangian computed with a coordinate section and using the null torsion and metricity conditions is the classical Einstein Lagrangian $\hat{L} (g,\p g)$. Of course the equations of motion are the same as in the section 4.1.

The asymptotic conditions should be coordinate independent of course so we need some asymptotic reference manifold, ideally a boundary.

Our purpose is to derive the superpotentials associated to $\hat{\L}$. We will analyse the conserved charges in the next section.

As for the previous Lagrangian \ref{actgrav}, only two of the three gauge symmetries will give non trivial Noether currents and associated superpotentials (as before, the $\kb$ local projective symmetry does not contribute and will be fixed in the Einstein gauge). The analogous equation to \ref{invar} for the Einstein Lagrangian $\hat{\L}$ (which now depends on the derivatives of $\th^a$ and $g^{ab}$ only) is 

\B   \d \hat{\L} - \ex \left(  \d \th^a  \we  \frac{\p \hat{\L}}{\p \ex \th^a} + \d g^{ab} \frac{\p \hat{\L}}{\p \ex g^{ab}} \right) \approx 0 \lb{eq:lagrvar2}\E

This formula can be used for both symmetries:

\VS

-  The $gl(D,\Re)$ symmetry: Due to the surface term added, the variation of the Einstein Lagrangian does not vanish anymore but is equal to a surface term $\d_\l \hat{\L} = \ex \left( \ex \l^a_{\ b} \we \sqrt{-g} g^{be} \S_{ae}\right)$. Note that this is in some sense a definition of $\d_\l \hat{\L}$ because the Noether method only gives quantities up to a exact form. If we use this in \ref{eq:lagrvar2}, the corresponding Noether current and superpotentials for $\hat{\L}$ are exactly the same as for $\L$, equation \ref{avoir} (and the definitions which follow it):

\VS

$\hat{\J}^b_{\ a} = \J^b_{\ a}$

$\hat{\U}^b_{\ a} = \U^b_{\ a} $

$\hat{\J}_\l = \J_\l$

$\hat{\U}_\l = \U_\l$

\VS

and the corresponding cascade equations are obviously the same.

\VS

- The diffeomorphism invariance: we use that under a Lie derivative the Einstein Lagrangian transforms as usual, i.e., $\d_\xi \hat{\L} = \ex \cdot \i_{\xib} \hat{\L}$ and that $\d_\xi \th^a = \li_{\xib} \th^a$, $\d_\xi g^{ab} = \li_{\xib} g^{ab} = \i_{\xib} \cdot \ex g^{ab}$, then formula \ref{eq:lagrvar2} becomes (compare equation \ref{totder})

\B  \ex \hat{\J}_\xi := \ex ( \xi^{\r} \hat{\t}_{\r} + \ex\xi^{\r} \we \hat{\s}_{\r}) \approx 0 \lb{totdere} \E

\noindent and direct computations show that

\VS
$ \hat{\t}_{\r} = \i_\r \hat{\L}  - \li_\r  \th^a \we \frac{\p \hat{\L}}{\p \ex \th^a} - \li_\r  g^{ab} \frac{\p \hat{\L}}{\p \ex g^{ab}}$

$\hspace{0.23in} = \t_{\r} + \ex \cdot \i_\r \Sg$

$\hspace{0.23in} = -\frac{\sqrt{-g}}{2k} \left( \w^c_{\ \r} \we \w^{ab} \we \S_{cab} + \w^{ac} \we \w_c^{\ b} \we \S_{\r ab}\right)  + \ex \i_\r \theta^a \we \left(-\frac{\sqrt{-g}}{2k} \w^{bc} \we \S_{bca} \right)$

$\hspace{0.5in} + \frac{1}{2k} \i_\r \left( \w^a_{\ e} \right) \we \cex (\sqrt{-g} g^{eb} \S_{ab}) + \frac{1}{2k}\i_\r \left( \th^a \right)\cex \left( \sqrt{-g} g^{dc} \S_{bca}\right) \we \w^b_{\ d}$

$ \hat{\s}_{\r} = - i_\r \th^a \ \frac{\p \hat{\L}}{\p \ex \th^a} $

$\hspace{0.23in} = \s_{\r} + \i_\r \Sg$

$\hspace{0.23in} =-\frac{\sqrt{-g}}{2k} \w^{ab} \we \S_{\r ab} $

\VS 

Note that $\hat{\J}_\xi =\J_\xi + \ex \left( \i_\xi \cdot \Sg \right)$ where $\J_\xi$ was given in \ref{totder}.

So all the cascade equations derived in the past sections can be used here by 
just hatting them and using these new formulas for practical computations. Note 
also that the theorem 2  of section 4.3 and its proof can be repeated 
identically.

\VS

The relation between $\t_\r$ and $\s_\r$ with the $gl(D, \Re)$ objects of Frauendiener \c{Fr} (see discussion which follow theorem 2 of section 4.3) is as follows:

Let us derive the cascade equations for the the $gl(D, \Re)$ vector $\xi^a$ defined as:

\VS

$\xi^a = \xi^\r \ \theta^a_\r$

\VS

where $\theta^a_\r$ will be fixed. Using this decomposition in \ref{totdere} we obtain that:

\B  \ex ( \xi^{a} \hat{\t}_{a} + \ex\xi^{a} \we \hat{\s}_{a}) \approx 0 \lb{totdereh} \E

where now

\VS

$ \hat{\t}_a =\hat{\t}_\r\ \theta^\r_a + \ex\theta^\r_a \we \s_\r $

$\hspace{0.23in} = -\frac{\sqrt{-g}}{2k} \left( \w^b_{\ a} \we \w^{cd} \we \S_{bcd} + \w^{bc} \we \w_c^{\ d} \we \S_{abd}\right) $

$\hspace{0.5in} + \frac{1}{2k} \omega^b_{ac}  \we \cex (\sqrt{-g} g^{cd} \S_{bd}) + \frac{1}{2k}\cex \left( \sqrt{-g} g^{dc} \S_{bca}\right) \we \w^b_{\ d}$

$ \hat{\s}_a = \hat{\s}_\r \theta^\r_a$

$\hspace{0.23in} =-\frac{1}{2k} \w^{bc} \we \S_{abc} $

\VS

What we did in the last equations is just a trick to change the$\ _\r$ index (curved manifold index) into a$\ _a$ index ($gl(D, \Re)$ index).

Similar objects have been encountered before in the works of Sparling, Nester-Witten, Dubois-Violette \& Madore and Frauendiener. As we already said, it is Frauendiener who finally defined their most general version, namely on the linear frame bundle $L(M)$ in $D$ dimensions:

\VS

$ \tilde{\t}_a = -\frac{\sqrt{-g}}{2k} \left( \w^b_{\ a} \we \w^{cd} \we \S_{bcd} + \w^{bc} \we \w_c^{\ d} \we \S_{abd}\right) $

$ \tilde{\s}_a = \hat{\s}_a$

\VS

It is easy to see that the $\s_a$'s (hatted and tilded) coincide and also the 
$\t_a$'s if we set the nonmetricity and the torsion to zero (more precisely the hatted quantities coincide with the pullback along an arbitrary section of the 
tilded ones).
What we have just achieved is to derive from a variational principal and 
symmetry arguments some objects whose existence appeared before rather mysterious.

\VS

Finally, we can define as in section 4.5 the total hatted Noether $\xi$-dependent superpotential, exactly in the same way as equation \ref{bigno}.

\VS

Let us now make contact with the ordinary second order formalism. We are not going to repeat here the complete analysis of section 3 for the Einstein Lagrangian
case:

\B  \hat{L} = \frac{1}{2k} \sqrt{-g} R -\p_\m S^\m =\frac{1}{2k}  \sqrt{-g} g^{\a\b} (\Gamma^\eta_{\a\d}\Gamma^\d_{\eta\b} - \Gamma^\eta_{\eta\d}\Gamma^\d_{\a\b}) \lb{he}\E

Where 

\B S^\m := \frac{1}{2k} \sqrt{-g} \left( \Gamma^\m_{\a\b}g^{\a\b} - \Gamma^\b_{\a\b}g^{\m\a} \right) \lb{surface} \E

Let us just give crucial differences and the principal results:

\VS

- The analogue of equation \ref{eqbase} is

\B \p_\m \hat{J}_\xi^\m \approx 0 \E

\B  \Leftrightarrow \p_\m (\xi^\r \hat{T}_\r ^{\ \m} + \p_\n \xi^\r \hat{U}_\r^{\ \m\n} +  \p_\d \p_\n \xi^\r \hat{V}_\r^{\ \m \n \d}) \approx 0 \lb{eqbasee} \E

Where 

\VS

$ \hat{T}^{\ \m}_\r := \d^{\ \m}_\r \hat{L} - \frac{\p \hat{L}}{\p \p_\m g_{\a\b}} \p_\r g_{\a\b} \lb{ein}$

$\hspace{.26in}            =\frac{\sqrt{-g}}{2k} \left[ \d^{\ \m}_\r (\G^\a_{\b\g}\G^\b_{\a\d}-\G^\a_{\a\b}\G^\b_{\g\d})g^{\g\d} +\G^\b_{\r\a}\G^\g_{\g\b}g^{\m\a}-\G^\b_{\r\b}\G^\g_{\g\a}g^{\m\a} \right.$

$\hspace{1.3in} \left. + \G^\a_{\r\a}\G^\m_{\b\g}g^{\b\g} + \G^\m_{\r\a}\G^\b_{\b\g}g^{\a\g} -2 \G^\m_{\a\b}\G^\a_{\r\g}g^{\b\g} \right] $

\VS

$\hat{U}_\r^{\ \m\n} := - \frac{\p \hat{L}}{\p \p_\m g_{\a\b}} (\d_\a^{\ \n} g_{\r \b} +\d_\b^{\ \n} g_{\r \a})$

$\hspace{.31in}            =\frac{\sqrt{-g}}{2k} \left[ \d^{\ \n}_\r ( \G^\m_{\a\b} g^{\a\b} - \G^\b_{\a\b} g^{\m\a}) + \d^{\ \m}_\r \G^\b_{\a\b} g^{\n\a} + \G^\a_{\a\r} g^{\m\n} -2 \G^\n_{\a\r} g^{\n\a} \right]$

\VS

$ \hat{V}_\r^{\ \m \n \d} = V_\r^{\ \m \n \d} = \frac{\sqrt{-g}}{2k} \left[ \frac{1}{2} g^{\d\m} \d_\r^{\ \n} + \frac{1}{2} g^{\n\m} \d_\r^{\ \d} -  g^{\d\n} \d_\r^{\ \m} \right] $

\VS

-The cascade equations are the same equations as \ref{2cas1}-\ref{2cas4} but hatted.

- $\hat{L}$ is now a Lagrangian which depends only on the metric and its first derivative. One then may ask where the $\hat{V}_\r^{\ \m \n \d}$ term comes from? The answer is quite simple and is that now $\hat{L}$ is not anymore a scalar. In fact, it is easy to show that its variation under a diffeomorphism induces an inhomogeneous surface term due to the non tensorial part of the $\G$'s. In other words, $\d_\xi \hat{L} = \p_\m ( \xi^\m \hat{L} ) - \frac{1}{2k}  \p_\m (\sqrt{-g} ( \p_\d \p_\n \xi^\m g^{\d\n} - \p_\d \p_\n \xi^\n g^{\m\d}))$. We note finally that $\hat{V}_\r^{\ \m \n \d}= V_\r^{\ \m \n \d}$.

- $ \hat{T}^{\ \m}_\r$ has been found by Einstein himself and is ususaly called the ``canonical energy momentum Einstein pseudotensor".

- The Einstein canonical spin complex $\hat{U}_\r^{\ \m\n}$ is not more 
antisymmetric than its Hilbert brother. Its associated antisymmetric quantity 
is what is called the Freud superpotential \c{vF} (analogous to the 
M\o ller one, equation \ref{luck}): 

\B  \ _F\hat{U}^{\ \m\n}_\r  := \hat{U}^{\ \m\n}_\r - \p_\d W^{\ \n[\m\d]}_\r \lb{luckh} \E

\B \hat{T}_\r^{\ \m} \approx \p_\n \ _F\hat{U}_\r^{\ \m \n} \E

Where,

\B  _F\hat{U}_\r^{\m\n} := \frac{1}{2k} \frac{1}{\sqrt{-g}} g_{\r\a} \p_\b(-g(g^{\m\a}g^{\n\b}-g^{\m\b}g^{\n\a})) \lb{vf} \E

Note that $ _F\hat{U}_\r^{\m\n} = _M U_\r^{\m\n} + S^\m \ \d^{\ \n}_\r - S^\n \ \d^{\ \m}_\r$.

The abelian cascade trick will again naturally give us a non antisymmetric $\xi^\r$ dependent superpotential $\hat{U}^{\ \m\n}_\xi$ defined exactly as equation \ref{esup} hatted. Its corresponding antisymmetric quantity is just a non background version of the Katz superpotential \c{Ka}.

\B  \ _{Ka}\hat{U}^{\ \m\n}_\xi := \hat{U}^{\ \m\n}_\xi - \p_\d \left( \xi^\r W^{\ \n[\m\d]}_\r \right) \lb{luck2h} \E

\B \hat{J}_\xi^{\ \m} \approx \p_\n \ _{Ka}\hat{U}^{\ \m\n}_\xi \E

where we have that:

\VS

\B \ _{Ka}\hat{U}^{\ \m\n}_\xi :=\ _F\hat{U}_\r^{\m\n} \xi^\r + g^{\m\a} \p_\a \xi^\n - g^{\n\a} \p_\a \xi^\m \lb{KatzK} \E

$\hspace{.8in} = \ _K U^{\ \m\n}_\xi +S^\m \ \xi^\n - S^\n \ \xi\m $

\VS

What we have just obtained is the Katz superpotential if we add the condition that only asymptotically cartesian coordinates can be used to compute the conserved charges. We will come back to this important point in the next subsection. 

\VS

- The contact between the $gl(D,\Re)$ objects and the above formulas has been partially established by Frauendiener \c{Fr} and Szabados \c{Sz}. Remember that their definitions of $\tilde{\t}_a$ and $\tilde{\s}_a$ coincide with our hatted quantities (with the null torsion and metricity conditions added for the $\hat{\t}_a$). What they have shown is that:

\VS

\hspace{0.5in} $\bullet$ The dual of the pullback of $\tilde{\s}_a$ (or $\hat{\s}_\r$) along an holonomic section (in this case the $_a$ and $_\r$ indices are indistinguishable) is the Freud superpotential.

\hspace{0.5in} $\bullet$ The dual of the pullback of $\tilde{\t}_a$ (or $\hat{\t}_\r$ if $\X^a=\M^{ab}=0$) along an holonomic section is the Einstein pseudotensor.

\hspace{0.5in} $\bullet$ The pullback of $\tilde{\s}_a$ (or $\hat{\s}_a$) along an orthonormal section is the Nester-Witten form.

\hspace{0.5in} $\bullet$ The pullback of $\tilde{\t}_a$ (or $\hat{\t}_a$ if $\X^a=\M^{ab}=0$) along an orthonormal section is the Sparling form.

\VS

- With no more work than the results found in section 4.5 and the above theorems 
we can easily complete the picture:

\VS

\hspace{0.5in} $\bullet$ The dual of $\hat{\U}_\xi$ along an holonomic section is the Katz superpotential.

\hspace{0.5in} $\bullet$ The dual of $\hat{\J}_\xi$ along an holonomic section is the total hatted Noether current $\hat{J}_\xi^\m$.

\VS

- Finally, the non antisymmetry of the Einstein canonical spin complex $\hat{U}^{\ \m\n}_\r$ is explained in exactly the same way as we did for $U^{\ \m\n}_\r$.

\VS

The conclusion of this subsection is that the superpotential associated to some theory depends strongly on the choice of boundary conditions. This is expected because the cascade equations imply that in the case of a gauge symmetry all the charges have to be computed on a ($D-2$) hypersurface. If we complicate the story and we use more general boundary conditions then a background  superpotential has to be subtracted from the above defined superpotentials (see Rosen \c{Ro}, Cornish \c{Co}, Katz \c{Ka}, Katz, Bi\u{c}\'{a}k and Lynden-Bell \c{KBL}, Chru\'{s}ciel \c{Ch}). This is part of the subject of the following subsection, we shall return to it in II.

\subsection{Physical charges}

We have just seen that specific boundary conditions correspond to specific surface terms in the action which naturally do not change the equations of motion. As we know, the result of adding this surface term is not so innocuous. In fact, the Lagrangian looses its explicit scalar form (as we just saw, $\hat{\L}$ transforms as a scalar plus an inhomogenous term). The major consequence of that is that the associated parameter dependent superpotentials are not anymore covariant. For instance although the Komar superpotential (equation \ref{kom}) is covariant the Katz one (equation \ref{luck2h}) is not. Unfortunately for the covariance it is the second one which is derivable with the asymptotic spatial Dirichlet boundary conditions for the metric (and so with the Kerr solution for example) as we will see in this section.

\VS

\underline{asymptotic conditions}

The addition of a surface term $\Sg$ to the scalar Lagrangian $\L$ (see equation \ref{Ein}) replaces the ``Hilbert variational principle condition" (equation \ref{varcon2}) by the vanishing of

\B   \int_{\p M}   \d \th^a  \we  \frac{\p \hat{\L}}{\p \ex \th^a} + \d g^{ab} \frac{\p \hat{\L}}{\p \ex g^{ab}}  \lb{einvar}\E

This condition is more satisfactory in the light of the discussion which followed equation \ref{varcon2}.

As in the Yang-Mills case (section 2.2), the purpose here is not to precisely solve this condition, say for example for an asymptotically flat boundary. We will just illustrate what is going on with a Dirichlet condition for the metric in the coordinate-Einstein gauge (i.e. $\th^a$ = $\ex x^a$ and $\La$ chosen so that the torsion and the metricity vanish). The Dirichlet solution for the vanishing of equation \ref{einvar} in that gauge is

\B \lim_{r \rightarrow \infty} \d \gb = 0 \E

Where $\gb = g_{ab}\ \ex x^a \otimes \ex x^b$. The solution to that is just as usual $\gb | _{\p M} = \bar{\gb}$, where $\bar{\gb}$ is a given asymptotic boundary metric.

Remember that in the coordinate-Einstein gauge the remaining gauge symmetry is a linear combination of a diffeomorphism $\li_\xi$ and a $gl(D,\Re)$ gauge rotation with parameter $\l^a_{\ b}=-\p_b \xi^a$ (section 4.4). 
The result of that is just the well known definition of the Lie derivative in a Riemannian manifold, which for simplicity will also be called $\li_\xi$. Now this symmetry has to preserve the above boundary condition, that means that $\li_\xi \gb | _{\p M} = 0$. We obtain just the asymptotic Killing condition:

\B \lim_{r \rightarrow \infty} \left( \xi^\r \p_\r g^{ab} - \p_\r \xi^a g^{\r b} - \p_\r \xi^b g^{a \r} \right) = 0 \lb{kill} \E

Again it is important to find all the $\xi^\r(x)$ that satisfy this condition. After that, a recipe can be given to obtain all the conserved physical charges of the theory due to the gauge symmetry. Before that we will just comment on the non covariance of $_{Ka} \hat{\U}_\xi$ and on the way Katz cured it.

\VS

\underline{The need of a background metric} :

The problem of non covariance is important for practical computations. For example only Cartesian coordinates were allowed to compute say the mass of the Schwarzschild black hole in the Freud superpotential formula (which is equal to the Katz superpotential for constant $\xi^\r$).

The way to remedy this is now well known and was introduced in the case where the asymptotic metric is flat by Rosen \c{Ro} and Cornish \c{Co}. Technically it consists in introducing a background metric $\bar{g}_{\m\n}$ in the theory and replaces every non covariant piece by a covariant one with respect to this metric. We will just refer the interested reader to the works of Katz \c{Ka}, Chru\'{s}ciel \c{Ch} and Katz,  Bi\u{c}\'{a}k and Lynden-Bell \c{KBL} for flat and more general backgrounds (for example an anti de Sitter space). For completeness, let us note that the introduction of a non dynamical background connection was proposed in \c{FF}.

The point is that this background metric is nothing other than the boundary conditions we used to solve the Dirichlet problem (see above). Thus the conserved charge makes sense in general only on the boundary of the manifold $\p M$, where its corresponding superpotential is well defined. The idea of using the background metric $\bar{g}_{\m\n}$ everywhere would allow to define the non covariant superpotential on $M$; however the physical meaning of these objects in general is not clear (as for example a quasilocal mass).

In some special cases, when there exists a $\mathit{global}$ space like Killing vector a quasilocal charge can be defined. We postpone the discussion of the angular moemntum to the last part of this subsection.

\VS
\underline{The conserved charges}

Suppose as in our general discussion (section 2.1) that a portion of spacetime 
(between two times, say $t_1$ and $t_2$) is bounded by a ($\mathit{D-1}$) timelike
 hypersurface at infinite distance $\Sigma_\infty$ and by two spacelike ($\mathit{D-1}$) 
hypersurfaces $\Sigma_1$ and $\Sigma_2$.  The possibility of having a black hole will not alter our discussion of the conserved charges, 
so we will ignore it. Without more computations, we give now the recipe to find all the conserved charges due to the local diffeomorphism invariance:

\VS

- Solve the Einstein equation with your chosen boundary conditions (we shall take the example of Dirichlet boundary conditions for the metric).

- Given these boundary conditions, derive the  parameter dependent Noether current $\J_\xi$ with its associated superpotential $\U_\xi$ (for example $_{Ka}\hat{\U}_\xi$ of equation \ref{luck2h} for the Dirichlet choice). If this superpotential is not covariant, use the appropriate background correction on the boundary of the manifold $M$.

- Find all the $\xi^\r(x)$ that satisfy the asymptotic Killing equation \ref{kill} say for the Dirichlet condition, or its analogue for the Neumann case, 
$$\lim_{r \rightarrow \infty} \d_\xi \w^a_{\ b} = 0 .$$
 They may form an infinite group.

- Integrate the equation $\J_\xi \approx \ex \U_\xi$ on the ($D-1$) surface $\Sigma_\infty$, bounded by $B_1$ at time $t_1$ and $B_2$ at time $t_2$ (as in our general discussion of section 2.1). The physical condition for charge conservation is that the charge does not ``run away", i.e. that $\J$ has to vanish on $\Sigma_\infty$. 
This implies in particular that the integral of $\U_\xi$ on $B_1$ will be the integral of $\U_\xi$ on $B_2$. In other words, the integral of $\U_\xi$ on a $\mathit{D-2}$ hypersurface (say now $B_\infty$) at infinite distance and fixed time $\mathit{is}$ the conserved quantity, and $\mathit{nothing}$ more. 
Warning: but for instance the integral of $\J_\xi$ over a fixed time spacelike hypersurface is not in general (for example when a black hole exists) conserved contrary to the usual Noether charge associated to a global symmetry.

- Compute for each $\xi^\r(x)$ the associated conserved charge given by:

\B Q(\xi^\r(x))=\int_{B\infty} \U_\xi \E

with the parameter dependent superpotential associated to your boundary choice.

- The number of conserved charges will be given by the number of finite Q's which are 
``not always zero" (``not always zero" means that for example even if the linear 
momentum of Schwarzschild black hole is zero, a finite boost transformation can give 
it a non null contribution; in that case the linear momentum counts as a 
charge).  
The number of really independent charges (not connected by a asymptotic gauge 
transformations) will be given by the number of Casimirs of this subgroup.

\VS

\underline{Some comments on the total Noether current}

We would like to conclude with some comments on the total Noether current $\J_\xi$, and its associated equation

\B \J_\xi=\ex\U_\xi \lb{asequ} \E 

In our discussion on conserved charges we just integrated this equation at $\Sigma_\infty$. The other obvious choice is to integrate it on a ($D-1$) space like hypersurface $\Sigma$, bounded say by $B_1$ and $B_2$, where $_1$ or $_2$ can stand for an asymptotic or black hole boundary, or any other finite distance. This is the problem of quasilocal charges.

However, we have to be very careful because to perform such an integration, the current and the superpotential must be defined everywhere and not only at infinity. Let us show some cases where this can be done:

\VS

- The quasilocal angular momentum associated to a global Killing vector, namely $\bigtriangledown ^{(\m} \xi^{\n)}|_{\Sigma} = 0$. Remember that the Katz superpotential can be written as the Komar (one half of the usual definition) superpotential wich is covariant plus a non covariant part (see equation \ref{KatzK}):

\B _{Ka}\hat{U}^{\m\n}_\xi = \ _K U^{\m\n}_\xi + S^\m \ \xi^\n - S^\n \ \xi^\m \lb{KK1}\E

One of the timelike normals to the boundary of the hypersurface $\Sigma$ is orthogonal to the Killing vector $\xi^\m$ (which is spacelike). If $\Sigma$ is 
chosen tangent to the Killing vector so that the second normal to its boundary is also orthogonal to $\xi^\m$, then the last two terms of \ref{KK1} vanish after integration over $\Sigma$.

In that case, the Katz superpotential becomes covariant and so can be defined at  any point of the manifold,  it becomes  the covariant Komar superpotential.

On the other hand, the total Noether current $J^\m_\xi$ (see definition \ref{tder}) becomes in the Killing case just $\xi^\m L$ (remember that $\d_{\xi} g_{\m\n} = 0$). When integrated over $\Sigma$ this term will also vanish (remember the orthogonality between $\xi^\m$ and the normal of $\Sigma$). What we obtained is just that

\B \int_{B_1}\ _K \U_{\xi} = \int_{B_2}\ _K \U_{\xi} \E

Which shows the quasilocal nature of the conserved quantity.

The angular momentum of an axi-symmetric spacetime obviously satisfies the above conditions. As was shown by Katz, the one half factor in front of the old
Komar expression is welcome because it gives the right (absolute) value for the angular momentum.

Note finally that the existence of a quasilocal conserved charge could have been expected from Kaluza-Klein arguments. In fact, the presence of a global spacelike Killing vector allows us to dimensionally reduce the gravitational Lagrangian along this direction. It is well known that when such a reduction occurs, some abelian gauge field with its associated quasilocal (electric type) charge appears. Thus the quasilocal charge is just the one corresponding to the abelian internal symmetry left after a Kaluza-Klein reduction.

\VS

- The proof of positivity of the gravitational mass is also an example where the equation \ref{asequ} has been used successfully \c{Wi} \c{Ne}. The integral of $\J_\xi$ on $\Sigma$ was shown to be positive and the integral of $\U_\xi$ on the black hole horizon to vanish \c{Gi}. 
This implied the positivity of the integral of $\U_\xi$ at spatial infinity. We would like to point out here that the superpotential used by Witten \c{Wi} (and later in a covariant way by Nester \c{Ne}) was explicitly covariant, spinor dependent and reduces to what is called the Nester-Witten form ($\hat{\s}_a$ of section 5.1) in the constant spinor case. It has been shown to come from  the asymptotic  local supersymmetry invariances of N=1 D=4 supergravity \c{Hu}.

\VS

- Finally another historical example where the equation \ref{asequ} has been used is in the proof of the first law of black hole thermodynamics \c{BCH}. The Killing vector is timelike but in vacuum $L \approx 0$ and so  $\J_\xi$ (Komar current) vanishes in the bulk of a spacelike hypersurface. This implies a relation between the integral of $\U_\xi$ at the horizon of the black hole and at spatial infinity wich was the first step in the proof of the first law. Note that in that case, the covariant superpotential used was just the original Komar superpotential (multiplied by its historical factor of two).  Some improved version of the first law was given by Wald \c{Wa}.

\VS

\section{Fluids: a case of constrained gauge parameter}

This section is meant to be readable independently of the previous three.

\subsection{The theory and the cascade equations associated to its $\mbox{\boldmath$\mathit{sdiff(V_a)}$}$ gauge symmetry}

In this last section we will discuss another case where the cascade equations can be 
useful to find conserved quantities. We will treat the case of non relativistic fluids,
(the relativistic case can be studied in a similar way, work in progress), in a 
Lagrangian formulation in contrast with the more usual Eulerian discussion. 

\VS

In this context, the basic fields of our theory are the fluid-particle Eulerian coordinates $x^i (a^a,\tau)$. The cells of fluid are labelled by the $a^a$ at a given time $\tau$. The $i,j,k...$ indices will be used for the laboratory space (called $x$-space) whereas the $a,b,c...$ will be for the internal label space (called $a$-space) both with same dimension $D$. The labelling follows the fluid particles along the dynamics. The labels are the Lagrangian coordinates. The domain of $a^a$ is a manifold $V_a$ without boundary (see paper II for boundaries).

The action to extremize is the integral of the following Lagrangian \c{Sa} over the $a$-space:

\B L = \frac{1}{2} \left( \frac{\p x^i}{\p \tau} \right)^2 - e \left( \det{\frac{\p x^i}{\p a^a}},\ s(a^a) \right) - \Phi (x^i) \lb{fluid}\E

$\Phi (x^i)$ is the potential of some external force.

Here $e$ is just the specific internal energy, a given thermodynamic function of $\det{\frac{\p x^i}{\p a^a}}$ and of the specific entropy $s$. The important hypothesis here is that the entropy $s$ depends on the labels but not on the time $\tau$, it is an adiabaticity or isentropy condition. 
We could have used the pressure or any other macroscopic thermodynamic variable. We will see in the following that the presence of such a conserved non uniform function breaks the maximal infinite dimensional $\tau$-independent relabelling gauge group allowed by the theory, namely $\mathit{SDiff(V_a)}$, the group of all diffeomorphisms that preserve the $D$-volume to (proper)subgroups. The amazing thing is that the number of mesurable physical charges will be dramatically altered by such reductions.

We will not analyse here the equations of motion of this Lagrangian and its relation with the more usual formulation of Euler equation. We just refer the interested reader to the excellent review \c{Sa}.

\VS

Another way to look at this theory is to say that it is the dynamics of the mapping 
$x^i(a^a,\tau)$ from the $a$-space onto the $x$-space ($a^a \rightarrow x^i$) as a
function of the  time $\tau$. We will suppose that this mapping is invertible. This will allow us to come back to the Eulerian description (i.e. velocities of the fluid as functions of the $x^i$'s) using the inverse formula $a^a=a^a(x^i,t)$ (see section 6.3).

The $a$-space contains of course a volume form to allow integration, its density has been normalised for convenience to one. Its pullback with the inverse $x$-map will then induce a volume form on $x$-space with density $\r$ which is given by the inverse of the Jacobian of the transformation $a^a \rightarrow x^i$:

\B \r = Det \left| \frac{\p x^i}{\p a^a} \right|^{-1} \E

The labels will be taken so that $\r$ gives correctly the mass density of the fluid. 

The $x$-space admits the Euclidean metric $\eta_{ij}$. To simplify our example this metric was assumed to be flat but in more general cases it could depend on the $x^i$ fields. This is the case of general relativistic fluids which can be analysed similarly. Its pullback along the $x$-map will induce a metric on the $a$-space, which will not be invariant, i.e.,

\B g_{ab} = \frac{\p x^i}{\p a^a} \frac{\p x^j}{\p a^b} \eta_{ij},\hspace{0.3in} \frac{\p g_{ab}}{\p \tau} \neq 0 \E

Let us first analyse the $\mathit{homentropic}$ case (barotropic if the pressure is taken as the given thermodynamic function) where $S(a^a) = S_0 = C^t$.

If we make a completely general (internal) coordinate transformation

\B \d a^a = \xi ^a (a^a) \lb{varf} \E

\noindent the Lagrangian \ref{fluid} will not in general vary by a total 
derivative; instead, 
$\d L = \xi^a \p_a L$. Something special happens in general 
relativity where the 
volume form is metric compatible of density $\sqrt{-g}$. The variation of this term 
provides the missing $\p_a \xi^a L$ part which allows to complete the total derivative,
in that case, the symmetry group contains  all the spatial diffeomorphisms of the internal coordinates denoted by $\mathit{Diff(V_a)}$. 
The absence of such an invariant metric in the $a$-space does not allow us to use that 
trick. However imposing on the gauge parameter to be divergenceless, $\p_a \xi ^a = 0$ 
the transformation \ref{varf} is now a gauge symmetry of the fluid Lagrangian \ref{fluid}, $\d L =  \p_a (\xi^a L)$. The infinite dimensional gauge group is then $\mathit{SDiff(V_a)}$, the group of $\tau$-independent diffeomorphisms which preserve the volume form.
 Let us insist that this has nothing to do with incompressibility which does not hold here, in fact the incompressible case will be discussed in paper II.
Under such a local transformation, the variation of the fluid field is given by $\d x^i (a^a,\tau) = \xi ^a \p_a x^i$. Let us now apply the cascade machinery to the Lagrangian \ref{fluid}, equations \ref{cur1} and \ref{cur11} give

\B \p_\a \left( T^{\ \a}_a \xi^a \right) \approx 0 \lb{noflu} \E

Where the $\a$ indices combine $(\tau, a^a)$ and $T^{\ \a}_a := \frac{\p L}{\p \p_\a x^i} \p_a x^i - \d^{\ \a}_a L$ is the canonical energy momentum tensor. The cascade equations associated to this symmetry are given by:

\B \p_\a \xi^a \ T^{\ \a}_a \approx 0 \lb{cas1f} \E

\B \xi^a \ \p_\a T^{\ \a}_a \approx 0 \lb{cas2f} \E

Since $\xi^a$ is not arbitrary but constrained by the volume preserving condition (and of course also time independent), the theorem 1 of the end of section 4.2 is not applicable any more. Thus we cannot say that the Noether current $T^{\ \a}_a$ identically vanishes due to the locality of the symmetry but instead that (see equation \ref{cas1f})

\VS

$$ \p_b \xi^a \  T^{\ b}_a = 0 \hspace{.3in} \Rightarrow \hspace {.3in} T^{\ b}_a = \d^{\ b}_a \Lambda $$

\VS

Where $\Lambda$ is a function (a Lagrange multiplier) which can be explicitly computed, $\Lambda = \frac{P}{\r} - L$, where $P$ is the pressure and is defined as usual by $P=-\frac{\p e}{\p \r^{-1}}$. Using the definition of $T^{\ b}_a$ we easily see that this implies that the Lagrangian \ref{fluid} can depend only on the determinant of $\frac{\p x^i}{\p a^a}$ (i.e. on the density of the fluid $\r$) . With that, equation \ref{cas2f} becomes:

\B\xi^a \hspace{.2in}  ( \p_\tau  A_a \approx - \p_a \Lambda) \lb{eckart} \E

\noindent where we defined $A_a := T^{\ \tau}_a$, which is just $\frac{\p x^i}{\p \tau} \frac{\p x^j}{\p a^a} \eta_{ij}$. For further discussion in the next subsection, note that this can be rewritten as

\B \p_\tau \left( A_a \xi^a \right) = - \p_a \left( \Lambda \xi^a \right) \lb{prif} \E

\noindent We will use this result to obtain the conserved charges of the theory.

But before that let us consider the case where the entropy is not anymore uniform. Even $\mathit{sdiff(V_a)}$ does not preserve the Lagrangian \ref{fluid}. In fact, its variation is given by  $\d L = -\frac{\p e}{\p s} \frac{\p s}{\p a^a} \xi^a$. So the gauge symmetry of the theory is reduced. In fact, we have to look for some $\xi ^a\ \frac{\p}{\p x^a}$ such that its Lie derivative on $s(a^a)$ vanishes, i.e.

 \B \xi^a \frac{\p s}{\p a^a} = 0 \lb{lent} \E

 This shows that we can only use the $\mathit{sdiff(V_a)}$ vectors which are tangent to the constant entropy $(D-1)$-dimensional hypersurfaces $W_a$. The entropy plays the role of a label and allows us to reduce the dimension of the problem with the associated partial breakdown of diffeomorphisms. The resulting gauge group is thus $\mathit{SDiff(W_a)}$. If there were more than one thermodynamic quantity in the Lagrangian which depended explicitly on the labels, say $p$ of them, then the gauge group would be $\mathit{SDiff}$ on each $D-p$ dimensional invariant set. The extreme case is when all the directions are broken, no more local symmetry remains, the system is then like frozen. 

This restriction of $\xi^a$ to be orthogonal to $\p_a s$ has important consequences on the cascade equation \ref {eckart}. In fact we can now only deduce that

\B \left( \p_\tau  A_a  + \p_a \Lambda \right)  \approx \l \p_a s  \lb{eckcon} \E

Where $\l$ is an arbitrary function, the Lagrange multiplier for $s$. 


All these observations have very important consequences on the number of 
conserved charges of our theory as we will see in the subsection 6.3.

\bigskip

\subsection{Conserved charges and forcing}

 We saw in the gravitational (or Yang-Mills) discussion that the important charge is 
the parameter dependent  (or gauge invariant) one. In simple words it just means that the physical charges that we will be able to measure in our laboratory are going to be scalars of the gauge group (so, quantities with no floating $\ ^a$ index and no $a^a$ dependence). 
Let us first integrate equation \ref{prif} over all the $a$-space with the 
 supposition that there is no boundary:

\B \p_\tau \int_{V_a} A_a \xi^a\ d^D a \lb{chaf} = 0\E

What we have to do now is just to find all the allowed $\xi^a$ as functions of the local fields that satisfy the volume preserving constraint and the $\tau$-independence. 
The key formula \ref{chaf} identifies the one dimensional subgroup of the infinite 
dimensional gauge group whose charge is computed. 

Let us say in words what remains to be done: the problem to find gauge invariant charges is now that of finding the diffeomorphism parameters $\xi^a $ that can be constructed 
in a tensorial way from the physical fields of the theory. This is done in 
detail  and full generality 
in the next section. Considering the importance of the helicity 
invariant of Moreau \c{Mor} and Moffatt \c{Mt}
in turbulence, especially in magnetohydrodynamics (dynamo problem, Kolmogorov cascade...) 
let us exhibit the diffeomorphism in label space that is responsible for helicity, in 
other words that is such that rigid diffeomorphisms along it have helicity as their
canonical Noether charge. Then we will give a general algorithm for impulsively 
modifying the charge (here the helicity) in a controlled fashion. We shall use 
components rather than differential forms to be read more widely despite some 
inevitable heaviness.

It turns out that the  vorticity (co)vector which can be expressed in 
Lagrangian coordinates $\omega^a:= (curl\, u)^i (\p a^a / \p x^i)$    
is the simplest possible $\xi^a$ one can think of.
It could have been guessed long ago that 
in order to increase the pseudoscalar density of helicity $u.\omega$ it is necessary to push (or 
kick) the fluid along its vorticity the resulting change in vorticity however 
might destroy the effect. In fact it does not, we shall presently show that pushing along the vorticity and proportionnally to it, that is only along vortex 
bodies is optimum. 

We would like now to propose as a general result the 

RULE: If there are a global or rigid Noether symmetry and 
 the associated charge  given with $x$ representing the field variables
by $\d x=\xi (x)$ and $Q= \int \J_\xi$ from eqs. (23) \ref{prif}, 
then the change of $Q$ under the impulsive forcing at some time t

$$\d x=0 \, , \, \d u=\xi (x)$$ 

\noindent is precisely equal to 
$$\d Q = (\p^2 L/\p u \p u). \xi .\xi$$
This is a positive quantity in view of the positivity of the acceptable 
kinetic terms. 

It is important to realise that the time independent symmetry variation
dictates the form of a time dependent kick (not a symmetry anymore) along 
itself that does increase the 
charge (with the proper sign). This is obvious for linear momentum and 
translation invariance but quite general. Note that  dropping the surface term 
is 
allowed for any spatial symmetry; for the energy or boost charges a separate 
analysis is required.
 
Let us now consider a simple example. To create
 helicity let us take a vortex ring with vorticity concentrated 
for definiteness on the core, helicity is zero by parity. Then let us force it 
to rotate along itself in effect creating a second vorticity distribution 
through it clearly this
is a situation of knotted vortex lines see \c{Mt} at the origin of helicity.

This can be adapted to 2 dimensions and leads to control of enstrophies. Let us recall that the first enstrophy is quite important in Kolmogorov's cascade.
We shall develop the applications of this general technique in further papers.
 
\subsection{Explicit ``abelian" symmetries}

The best way to do that is to rewrite everything in differential form notation
and we shall follow the review \c{AK}:

\VS

- First remember that there is no available invariant metric in the $a$-space and that the Hodge dual operation is not allowed.

- There exists a ($\mathit{D-1}$)-form associated to $\xi^a$, namely $\xib = \i_\xi \mb= \frac{1}{(D-1)\!} \e_{ab_2 \ldots b_D} \xi^a\ \ex a^{b_2}\we \ldots\we \ex a^{b_D}$, where $\mb$ is the volume $D$-form of the $a$-space. The volume preserving condition is then just the closure of $\xib$, $\ex \xib = 0$, or locally, $\xib =\ex \nb$, where $\nb$ is some ($\mathit{D-2}$)-form to be determined. If there exists some non uniform thermodynamic function, say for instance the entropy $s$, then $\xib$ must satisfy the additional constaint $\xib \we \ex s =0$, see equation \ref{lent}.

- $A_a$ will now be written as a $1$-form $\Ab := A_a \ex x^a$. So equation \ref{eckart} becomes $\p_\tau \Ab \approx - \ex \Lambda$ which implies that $\p_\tau \ex \Ab \approx 0$ (this is nothing but the Eckart law \c{Ec} derived also with Noether arguments by Salmon \c{Sa}). In the nonhomentropic case, equation \ref{eckcon} becomes $\p_\tau \ex \Ab =\ex \l \we \ex s$. Thus the supposition that the entropy is time independent implies that the conserved quantity will be now $\p_\tau \ex \Ab \we \ex s =0$.


-The conserved charge \ref{chaf} is in this notation $Q(\xib) = \int \Ab \we \xib$, with the extra condition (the $\tau$ independence of $\xib$) that $\p_\tau \xib =0$.

\VS

In summary, the problem of finding a conserved charge has been reduced to the problem of finding all ($\mathit{D-1}$)-forms $\xib$ which satisfy

\VS

\hspace{.3in} - Closure, $\ex \xib = 0$.

\hspace{.3in} - Time independence, $\p_\tau \xib = 0$.

\hspace{.3in} - preservation of the thermodynamic function(s), if any, $\xib \we \ex s =0$.

\VS

To solve first the time independence condition we can observe that if the only available field of our theory is $\Ab$ then any $\tau$ independent local function (or form) has to depend only on $\ex \Ab$ in the homentropic case and on $\ex \Ab \we \ex s$ in the nonhomentropic one. $\ex \Ab$ is just the Lagrangian version of vorticity. So, the local, time-conserved,

\VS

\hspace{.3in} $\bullet$ \underline{functions} are 

- the entropy or any other thermodynamical given function, if any. 

- only in the $\mathit{even}$ dimensional homentropic case ($D=2n$, $n \geq 1$) , $f_0=\frac{\ex \Ab \we \ldots \we \ex \Ab}{\mb}$ ($n$ number of $\ex \Ab$). 

- only in the $\mathit{odd}$ dimensional nonhomentropic case ($D=2n+1$, $n \geq 1$), $g_0=\frac{\ex s \we \ex \Ab \we \ldots \we \ex \Ab}{\mb}$ ($n$ number of $\ex \Ab$). 

Of course, any function of these functions is again a $\tau$ conserved function. Note in particular that in the homentropic (uniform entropy) $\mathit{odd}$ dimensional case, there is simply no $\tau$ conserved local function.

\VS

\hspace{.3in} $\bullet$ \underline{$1$-forms} are $\ex f$, where $f$ is any of the above time conserved functions. In fact, the only other one form we have at our disposal is $\Ab$ but it does not satisfy $\p_\tau \Ab =0$. Again, any conserved function times one of these one forms is also conserved.

\VS 

\hspace{.3in} $\bullet$ \underline{$2$-forms} are any wedge product of any two of the above $1$-forms and $\ex \Ab $ in the homentropic case.

\VS

\hspace{.3in} $\bullet$ \underline{$3$-forms} are any wedge product of the above $1$-forms and $2$-forms plus $\ex \Ab \we \ex s$ in the case of nonhomentropic fluid. 

\VS

\hspace{.3in} $\bullet$ \underline{$p$-forms}, $p\ \geq 4$ any wedge product of the above forms. 

\VS

Let us show how the analysis continues for specific cases:

\VS

\hspace{.3in} $\bullet$ The (homentropic) odd dimensional case, $D=2n+1$, $n \geq 1$: We are in the case where no conserved function exist. From the above discussion, we can see that only even conserved forms exist. Fortunately $\xib$ should be an even closed form (of degree $2n$). Then the only local possibility is $\xib =\ex \Ab \we \ldots \we \ex \Ab$ ($n$ times).
The conserved charge is thus, following equation \ref{chaf}:

\B Q = \int_{V_a} \Ab \we \left( \ex \Ab \we \ldots \we \ex \Ab \right) \lb{helicity} \E

The Eulerian version of these results already exists in the litterature 
\c{TS} \c{KC} and are just generalisations of the  helicity. The relation will be given in the last subsection.

\VS

\hspace{.3in} $\bullet$ The (homentropic) even dimensional case, $D=2n$, $n \geq 1$: The most important difference with the above case is that now we have an infinite number of $\tau$-conserved functions, namely $f_0$ and any function of it. However, the only non vanishing conserved $2$-form is $\ex \Ab$ because if f and g are two arbitrary functions of $f_0$, then $\ex f \we \ex g = \frac{\p f}{\p f_0} \frac{\p g}{\p f_0} \ex f_0 \we \ex f_0 =0$. 
We then can look for all the odd ($2n-1$)-dimensional closed forms $\xib$ which can be constructed from the above ingredients. It is not difficult to see that the most general possibility is $\xib=\ex f \we \ex \Ab \we \ldots \we \ex \Ab$ ($n-1$ times), where again $f$ is $\mathit{any}$ function of $f_0$. Then the conserved charges are 

\B Q_f = \int_{V_a} \Ab \we \left( \ex f \we \ex \Ab \we \ldots \we \ex \Ab \right) \lb{enstrophy} \E

Note that if we integrate by parts, the above definition is nothing but $Q_{\bar{f}} = \int_{V_a} f .f_0 \mb =\int_{V_a} \bar{f} \mb$, where $\bar{f}$ is another function of $f_0$ given by $f.f_0$. Since $f$ is arbitrary, $\bar{f}$ is arbitrary too. The number of charges is infinite. 

There is an intuitive way to understand the meaning of this infinity. If we take a Dirac $\d$ function for $\bar{f}$ (say $\d ( f_0 - C)$ where $C$ is an arbitrary constant) the conserved charges $Q_{\bar{f}}$ guarantee that each surface of constant $f_0$ is conserved. The fluid looks like an infinite number of ($D-1$)-dimensional subsystems at each slice of constant $f_0$. Then the arbitrary 
function $\bar{f}$ gives the relative weight to assign to each hypersurface of constant $f_0$ when we compute a conserved charge as an integral over the whole $D$-dimensional label space.

An Eulerian translation of this result (which is a generalisation of what is called enstrophy) will be given in section 6.3.

\VS 
\hspace{.3in} $\bullet$ The (nonhomentropic) odd dimensional case, $D=2n+1$, $n \geq 1$: Now we have an extra data which is a $\tau$ conserved function, namely the specific entropy s (again, the same is true for any other macroscopic thermodynamic function). The most important consequence of that is that now we can construct the analogue of $f_0$, namely $g_0 = \frac{\ex s \we \ex \Ab \ldots \we \ex \Ab}{\mb}$, which is time conserved.
With this extra data, the most general $2n$ closed form $\xib$ which in addition satisfies the symmetry constraint $\xib \we \ex s =0$ is simply $\xib= r(g_0,s)\ \ex g_0 \we \ex s \we \ex \Ab \ldots \we \ex \Ab$ ($n-1$ times), where $r$ is $\mathit{any}$ function of $g_0$ and s. The doubly infinite number of conserved charges are then

\B Q_{r} = \int_{V_a} \Ab \we \left( r\ \ex g_0 \we \ex s \we \ex \Ab \we \ldots \we \ex \Ab \right) \lb{enstroc} \E

This can be rewritten more simply after an integration by part as $Q_{\bar{r}} = \int_{V_a}  \bar{r} \mb$, where $\bar{r} := \left[ \int^{g_0} r(\tilde{g},s) d \tilde{g} \right].g_0$ is any function of $g_0$ and $s$. We will see in section 6.3 that the Eulerian version of $g_0$ is just what is usualy called the potential vorticity of Ertel \c{Er} in the three dimensional case.

The amazing point here is that in an odd dimensional space, the presence of a non uniform thermodynamic variable breaks the maximal symmetry group (namely $\mathit{SDiff(V_a)}$) with only one conserved invariant charge to a proper subgroup ($\mathit{sdiff(W_a)}$) (where $W_a$ is again a ($D-1$)-dimensional manifold) 
with an infinite number of conserved charges again invariant under relabeling. 

Note that the last formula \ref{enstroc} depends on the choice of the two parameter function $\bar{r}$. What really happens is that the presence of a non uniform entropy breaks our odd $D$-dimensional fluid theory into an infinite number of even $\mathit{(D-1)}$-dimensional ones. Take the example of a three dimensional fluid. 
When a gradient of entropy exists we can define a bidimensional theory on each slice of constant entropy. As we saw in the previous example, each of these bidimensional systems can again be interpreted as an infinite number of one dimensional systems, each of constant $g_0$. The $\bar{r}$ function gives the relative 
weight we assign to each one dimensional system when we compute some conserved charge.

As we can see in the following the same happens from even to odd dimentions.
Of course all the above equations and definitions can be translated without major work to the Eulerian formalism.

\VS

\hspace{.3in} $\bullet$ The (nonhomentropic) even dimensional case, $D=2n$, $n \geq 1$: Now we cannot use the function $f_0$ to construct conserved quantities because it is not any more $\tau$-conserved (remember that now $\p_\tau \ex \Ab = \ex \a \we \ex s$). 

The only time conserved function that we can use to construct $\xib$ is thus $s$. We again have to look for all gauge parameter $\xib$ ($\mathit{D-1}$)-forms that satisfy the $\tau$ independence constraint ($\p_\tau \xib = 0$), the divergenceless constraint ($\ex \xib = 0$) and the constraint $\xib \we \ex s = 0$. The result is that the only solution is $\xib=\ex F \we \ex \Ab \we \ldots \we \ex \Ab$ (where the number of $\ex \Ab$ factors is just ($\mathit{n-1}$) and F is an arbitrary function of the entropy $s$). The associated charge is

\B Q_F = \int_{V_a} \Ab \we \left( \ex F \we \ex \Ab \we \ldots \we \ex \Ab \right) \lb{enstroc2} \E

Again we find a single infinity of charges. The arbitrariness of the function 
$F$ has a similar meaning as the arbitrariness of the $\bar{r}$ function of the previous example. If for example we look for a four dimensional nonhomentropic fluid, then what really happens is that the theory is just an infinite number of three dimensional theories living in some slice of constant entropy. We can then compute the helicity for each of them which will be independently conserved. The total charge \ref{enstroc2}
is just each of these helicities weighted by an arbitrary function (for example a Dirac $\d$ function if we just want to look for the helicity of a specific slice). By analogy with the preceding example we will call these quantities the potential helicities.

\VS

In the case where we have more than one macroscopic thermodynamic function in our theory, the analysis, is completely straightforward if we follow the above examples. In fact the rule seems to be that if we have $p$ thermodynamic variables the gauge group is broken from $\mathit{SDiff(V_a)}$ to $\mathit{SDiff(W_a)}$ ($W_a$ is a ($D-p$)-dimensional manifold), with an infinite number of conserved charges if the effective dimension $\mathit{D-p}$ is even and just one in the odd case for each homogeneous submanifold of fluid (with uniform thermodynamic variable).

It is remarkable that the same group of volume preserving diffeomorphisms is 
absolutely essential in membrane theory and its application to define the 
mysterious M-theory by discretizing the surface and the group is under
intensive study. It is important to note that all the even dimensional flows 
share this feature of an infinite number of invariants, it is not a 
particularity of two dimensions. This may be encouraging for M-theory that 
requires us absolutely to consider less familiar objects than strings or even 
membranes.

\subsection{The Eulerian framework}

The purpose of this last section is just to give the translation of the above results in the more usual Eulerian language. For that we just need to do an ordinary change of variables in the integral formulas between the $a^a$ variables and the $x^i$ variables.

Before starting with some precise example, let us remember the general points:

\VS
- The change of variables will be $a^a \rightarrow a^a(x^i, t)$, and $\tau=t$. $t$ is just the Eulerian time, $\tau$ is the material or Lagrangian time. A conserved charge has to be invariant under the $\frac{\p}{\p \tau}$ derivative, which is in Euler time just the total derivative $\frac{D}{D t} := \frac{\p}{\p t} + u^i \frac{\p}{\p x^i}$. Of course $u^i$ is the Eulerian velocity.

- In the begining we fixed the volume form of the $a$-space to one. Now, the corresponding volume form in the Euler framework is just the inverse of the Jacobian of the transformation $a \rightarrow x$. We then must be careful to use Levi-Civita tensor densities and in the following the $\e^{ijk...}$ or $\e^{abc...}$ symbols will be always constant. The density $\r$ will then appear explicitly. 

\VS

We are now ready to translate all the results found in section 6.2 and 6.3:

$\bullet$ The total charge, equation \ref{chaf} becomes

\B \p_\tau \left[ Q(\xi) = \int_{V_x} u_i \xi^i\ \r\  d^D x \right] = 0 \lb{chafe} \E

Where $u_i = A_a \frac{\p a^a}{\p x^i}$ is the Eulerian velocity and $\xi^i = \xi^a \frac{\p x^i}{\p a^a}$.


\VS

$\bullet$ The helicity, equation \ref{helicity} becomes

\B Q= \int_{V_x} \e^{ij_1 k_1 \ldots j_n k_n} u_i \p_{j_1} u_{k_1} \ldots \p_{j_n} u_{k_n}\  d^{2n+1} x \E

This charge was found by other methods by Tartar \& Serre \c{TS} and by Khesin \& Chekanov \c{KC} as a $D$-dimensional generalisation of the usual helicity  \c{Mor} \c{Mt}.

\VS

$\bullet$ The enstrophies, equation \ref{enstrophy}, become

\B Q_{\bar{f}}= \int_{V_x}  \bar{f}(f_0)\ \r\  d^{2n} x \E

Where $f_0 = \r^{-1}\ \e^{i_1 j_1 \ldots i_n j_n}\ \p_{i_1} u_{j_1} \ldots \p_{i_n} u_{j_n}$ and $\bar{f}(f_0)$ is of course an arbitrary function of $f_0$. 

Again this result can be found in the work of Serre \c{TS} and  Khesin \& Chekanov \c{KC}.

\VS

$\bullet$ The potential vorticities, equation \ref{enstroc} becomes

\B Q_{\bar{r}}= \int_{V_x} \bar{r} (s , g_0)\  \r\  d^{2n+1} x \E

Where $g_0 = \r^{-1}\ \e^{ij_1 k_1 \ldots j_n k_n}\  \p_i s\  \p_{j_1} u_{k_1} \ldots \p_{j_n} u_{k_n}$ was already known in three dimensions as the potential 
vorticity \c{Er}, see for example \c{Sa}. $\bar{r}$ is an arbitrary function of 
$g_0$ and $s$. Remember that the double infinity of conserved charges was 
explained in section 6.3. It is just an infinite number of ($\mathit{D-2}$)-dimensional subsystems of constant entropy $s$ and constant potential vorticity $g_0$ weighted by $\bar{r}$.

\VS

$\bullet$ The potential helicity, equation \ref{enstroc2} becomes

\B Q_F= \int_{V_x} \e^{ijk_1 l_1 \ldots k_n l_n}\p_i F u_j \p_{k_1} u_{l_1} \ldots \p_{k_n} u_{l_n}\  d^{2n+1} x \E

Where $F$ is an arbitrary function of $s$. Here, the number of charges is just infinite. The reason is that our fluid, as in the previous example can be viewed as an infinite number of ($\mathit{D-1}$)-dimensional subsystems of constant entropy weighted by $F(s)$, whose odd dimensional character implies the existence of only one conserved charge, the helicity.

In conclusion let us comment on the impressive generality of the Noether 
approach, it did in fact impress her contemporaries. A poll among colleagues has
shown  that very few did actually know the full contents of the paper, and even 
fewer read it. 

\bigskip

{\bf Acknowledgments.}
\bigskip

Beyond most of the (living) authors quoted below we would like to thank 
M. Brachet, Y. Brenier, B. Carter,
E. Corrigan,
T. Damour, 
M. Dubois-Violette, S. Fauve, G. Gibbons, F. Hehl, M. Herzlich, 
E. Kuznetzov, D. Lynden-Bell, 
L. Mason, A. Pumir, A. Shnirelman and especially 
M. Henneaux. BJ is also grateful to the EU TMR contract ERBFMRXCT960012 for travel support, 
to the Newton Institute for a stimulating fortnight among fluid mechanics and gravity experts 
and to the A. von Humboldt foundation for an enjoyable visit to Bonn's MPIM and University. 
We presented in seminars at Cambridge, Oxford, Foljuif, Paris and Meudon various preliminary 
versions of the present results.

\end{document}